\documentstyle[12pt,psfig]{article}
\catcode`\@=11
\def\marginnote#1{}
\newcount\hour
\newcount\minute
\newtoks\amorpm
\hour=\time\divide\hour by60
\minute=\time{\multiply\hour by60 \global\advance\minute by-
\hour}
\edef\standardtime{{\ifnum\hour<12 \global\amorpm={am}%
    \else\global\amorpm={pm}\advance\hour by-12 \fi
    \ifnum\hour=0 \hour=12 \fi
    \number\hour:\ifnum\minute<100\fi\number\minute\the\amorpm}}
\edef\militarytime{\number\hour:\ifnum\minute<100\fi\number\minute}
\def\draftlabel#1{{\@bsphack\if@filesw {\let\thepage\relax
  \xdef\@gtempa{\write\@auxout{\string
    \newlabel{#1}{{\@currentlabel}{\thepage}}}}}\@gtempa
    \if@nobreak \ifvmode\nobreak\fi\fi\fi\@esphack}
     \gdef\@eqnlabel{#1}}
\def\@eqnlabel{}
\def\@vacuum{}
\def\draftmarginnote#1{\marginpar{\raggedright\scriptsize\tt#1}}
\def\draft{\oddsidemargin -.5truein
        \def\@oddfoot{\sl preliminary draft \hfil
        \rm\thepage\hfil\sl\today\quad\militarytime}
        \let\@evenfoot\@oddfoot \overfullrule 3pt
        \let\label=\draftlabel
        \let\marginnote=\draftmarginnote

\def\@eqnnum{(\theequation)\rlap{\kern\marginparsep\tt\@eqnlabel}%
\global\let\@eqnlabel\@vacuum}  }
\def\preprint{\twocolumn\sloppy\flushbottom\parindent 1em
        \leftmargini 2em\leftmarginv .5em\leftmarginvi .5em
        \oddsidemargin -.5in    \evensidemargin -.5in
        \columnsep 15mm \footheight 0pt
        \textwidth 250mmin      \topmargin  -.4in
        \headheight 12pt \topskip .4in
        \textheight 175mm
        \footskip 0pt

\def\@oddhead{\thepage\hfil\addtocounter{page}{1}\thepage}
        \let\@evenhead\@oddhead \def\@oddfoot{} \def\@evenfoot{}
}
\def\titlepage{\@restonecolfalse\if@twocolumn\@restonecoltrue\onecolumn
     \else \newpage \fi \thispagestyle{empty}\c@page\z@
        \def\thefootnote{\fnsymbol{footnote}} }
\def\endtitlepage{\if@restonecol\twocolumn \else  \fi
        \def\thefootnote{\arabic{footnote}}
        \setcounter{footnote}{0}}  
\catcode`@=12
\relax
\def\be{\begin{equation}}
\def\ee{\end{equation}}
\def\bea{\begin{eqnarray}}
\def\eea{\end{eqnarray}}
\def\simlt{\stackrel{<}{{}_\sim}}
\def\simgt{\stackrel{>}{{}_\sim}}

\newcommand{\beq}{\vspace{2mm}\begin{eqnarray}}
\newcommand{\eeq}{\end{eqnarray}\vspace{2mm}}

\newcommand{\nn}{\nonumber}

\newcommand{\ovm}{{\overline m}}

\relax

\begin{document}
\textheight 23cm
\textwidth 15.5cm
\topmargin-2.5cm

\begin{titlepage}
\begin{flushright}
CERN-TH/96-241\\
TUM-HEP-260/96\\
SFB--375/126\\
DESY 96-232\\
hep--ph/9612261 \\
\end{flushright}
\vskip 0.3in

\begin{center}
{\Large\bf Bottom-Up Approach and 

Supersymmetry Breaking}\\

\vskip .2 in
{\bf M. Carena}$^{\ddag,+}$,
{\bf P. Chankowski}$^{\ddag\dag}$\footnote{Work 
                    supported in part by the Polish Committee for 
                    Scientific Research and the European Union grant
                    ``Flavourdynamics'' CIPD-CT94-0034.}, 
{\bf M. Olechowski}$^{\S}$\footnote{On leave of absence from the
Institute of Theoretical Physics, Warsaw University,  Warsaw, Poland.}$^{*}$\\
{\bf S. Pokorski}$^{\S\S *}$ and 
{\bf C.E.M. Wagner}$^{\ddag}$ \vskip.4in

$^{\ddag}$CERN, TH Division, CH--1211 Geneva 23, Switzerland
\vskip .1 in

$^+$Deutsches Elektronen Synchrotron, DESY, D--22603 Hamburg, Germany
\vskip .1 in

$^{\S}$Institut f\"ur Theoretische Physik, 
Physics Department T30, \\
Technische Universit\"at M\"unchen, D$-$85748 Garching, Germany
\vskip .1 in

$^{\S\S}$ Institute of Theoretical Physics, Warsaw University, Warsaw, Poland\\
and Max-Planck-Institut f\"ur Physik, Werner-Heisenberg-Institut,\\
F\"ohringer Ring 6, D--80805 Munich, Germany
\end{center}
\vskip .6cm

\begin{center}
{\bf Abstract}
\end{center}
\begin{quote}
We present a bottom-up approach to the question of supersymmetry breaking in 
the MSSM. Starting with the experimentally measurable low-energy 
supersymmetry-breaking 
parameters, which can take any values consistent with present 
experimental constraints, we evolve them up to an arbitrary high energy scale.
Approximate analytical expressions for such an evolution, valid for low and 
moderate values of $\tan\beta$, are presented. 
We then discuss qualitative properties of the high-energy parameter space
and, in particular, identify the conditions on the low energy spectrum that 
are necessary for the parameters at the high energy scale to satisfy simple 
regular pattern such as universality or partial universality.
As an illustrative example, we take low 
energy parameters for which light sparticles, within the reach of the LEP2 
collider, appear in the spectrum, and which do not affect 
the Standard Model agreement with the precision measurement data.  
Comparison between supersymmetry breaking at the GUT scale and at a low energy
scale is made.
\end{quote}
\vskip.8cm

\end{titlepage}

\setcounter{footnote}{0}
\setcounter{page}{0}
\newpage

\noindent
\textheight 23.5 cm
\textwidth 15 cm
\voffset -1 cm
\hoffset -0.5 cm
\topmargin -0.0 cm

{\bf 1. INTRODUCTION}
\vskip 0.3cm

In supersymmetric theories the mechanism of supersymmetry breaking
remains a fundamental open problem. Its low energy manifestation is
the supersymmetric spectrum. Therefore, one may hope to get some
insight into this problem from experiment. Currently, the most popular
view on supersymmetry breaking is that the parameters
of the low-energy effective theory have their origin in the GUT (or
string) scale physics. Supersymmetry is spontaneously broken in an
invisible sector, and this effect is transferred to our sector through 
supergravity interactions at the scale ~$M_{Pl}$ ~(see e.g. \cite{NILLES}).
Other models  have also been proposed \cite{DINE}, in which the supersymmetry 
breaking is comunicated to the electroweak sector through some other
messengers at the energy scale ~$M\ll M_{Pl}$. ~
In the Minimal Supersymmetric Standard Model (MSSM), the soft supersymmetry
breaking parameters at the large scale 
\footnote{The evolution of the soft supersymmetry breaking
          parameters from ~$M_{Pl}$ ~to ~$M_{GUT}$ is strongly model dependent
          and hence will not be discussed in the present article.} 
are connected to their low energy values via the renormalization group
equations (RGEs), which do not contain any new unknown parameters.
Given a simple theoretical {\sl Ansatz} for the pattern of soft supersymmetry
breaking terms at the scale where supersymmetry breaking is transmitted
to the observable sector, one can study superpartner spectra in the top-down
approach. It is clear that any particular {\sl Ansatz} for parameters 
at the high energy scale selects only a small subset of the whole 
low energy parameter space. In order to have a broad overview of the 
low-energy$-$high-energy parameter mapping,
it is of interest to supplement the top-down approach
with a bottom-up one, to learn how certain qualitative features
of the low energy spectra reflect themselves on the pattern of soft 
supersymmetry breaking terms at different energy scales. Of course, 
the direct measurement of the superpartner and Higgs boson spectra
and various mixing angles in the sfermion sector will permit (in the
framework of MSSM) a complete bottom-up mapping. This will determine
the pattern of the soft parameters at any hypothetical scale ~$M$ ~
of supersymmetry breaking and will have a major impact on our ideas
on its origin.

In this work we discuss the bottom-up mapping for the set of
parameters ~$\mu$, ~$M_2$, ~$m^2_{H_1}$, ~$m^2_{H_2}$, ~
$B\mu$ ~and the third-generation squark mass
parameters ~$m^2_Q$, ~$m^2_U$, ~$m^2_D$, ~$A_t$ ~and ~$A_b$. ~To a very 
good approximation, this is a closed set of parameters, whose RG running 
decouples from the remaining parameters of the model
\footnote{In the small or moderate $\tan\beta$ regime, 
          the dependence of the RG running of these parameters
          on slepton (and the two first generation sfermion) masses
          comes only through the  small hypercharge {\it D}-term contributions;
          see eqs. (\ref{eqn:m2k0}$-$\ref{eqn:dterm}).}.
We shall concentrate on the region of small to moderate values
of ~$\tan\beta$ ~in  which, for ~$m_t = 175 \pm6$ GeV, ~the bottom quark 
Yukawa coupling effects may be neglected.
For this case we present analytic expressions (at the one loop level) for 
the values of our set at ~$M_{GUT}$ ~and at any ~$M < M_{GUT}$ ~in terms of 
its values at the scale ~$M_Z$. ~The equations are valid for arbitrary 
boundary values of the parameters at the scale ~$M$. ~

In the absence, as yet, of any direct experimental measurement of the low 
energy parameters, we discuss the mapping of the low energy region, which 
is consistent with all existing experimental constraints and is of interest 
for LEP2.
Although indirect, the precision data provide  very relevant information
on the phenomenologically acceptable supersymmetric parameters. A few per
mille accuracy in the agreement of the data with the Standard Model (SM)
\cite{LEPEWWG}
is a very strong constraint on its extensions. The best fit in the SM
gives some preference to a light Higgs boson ~($M_h=145^{+160}_{-80}$ GeV ~
at ~$1\sigma$ \cite{LEPEWWG,MY_MH}) ~in agreement with the predictions of 
the MSSM. Thus the MSSM, with the superpartners sufficiently heavy to decouple
from the electroweak observables, gives a global fit to the electroweak data 
as good as the SM \cite{ABC,ELLIS,MY_MSSM}. A more careful study of the 
electroweak observables shows that a quantification of 
\lq\lq sufficiently heavy" strongly depends on the type of superpartner. 
Indeed, most of the electroweak observables are sensitive mainly to additional
sources of custodial ~$SU_V(2)$ ~breaking in the 
\lq\lq oblique" corrections. Thus, 
they constrain left-handed squarks of the third generation to be heavier than 
about ~$300-400$ GeV, ~but leave room for other much lighter (even lighter 
than ~$M_Z$) ~superpartners \cite{MY_MSSM}. 
This high degree of screening follows from the basic structure of the MSSM. 
Thus, sleptons, particles in the gaugino/higgsino sector, and even the 
right-handed stop, can be lighter than ~$Z^0$. ~One remarkable exception (and 
the only one among the ~$Z^0$-pole ones) is the value of ~
$R_b\equiv\Gamma(Z^0\rightarrow b\overline b)/\Gamma(Z^0\rightarrow hadr)$ ~
which is sensitive precisely to some of the masses just mentioned. 
In the MSSM with chargino and stop masses close 
to ~$M_Z$, ~values of ~$R_b$ ~ranging from ~0.2158 ~(the SM prediction)
up to ~0.218 (0.219) ~for small (large) ~$\tan\beta$ 
\footnote{For large ~$\tan\beta$, ~also ~$A^0$ ~has to be light.} 
can be naturally obtained depending on the chargino 
composition ~($r\equiv M_2/|\mu|$) ~and the stop mixing angle
\cite{BF,KANE,SOLA,MY_MSSM,MY_RB}.
For low values of ~$\tan\beta$, ~maximal values of ~$R_b$ ~are obtained  
for negative ~$\mu$ ~and ~$M_2\approx -\mu$ ~\cite{MY_RB,KANE1,DPROY}.
Although the recent data \cite{LEPEWWG} still prefer values of ~
$R_b$ ~slightly larger than predicted in the SM, this preference is 
not very strong statistically.
Superpartners with masses around (or even smaller than) the ~$Z^0$ ~mass
are, of course, also constrained by limits from direct searches, Higgs boson
mass limit and the ~$b\rightarrow s\gamma$ ~decay. The recent completion of
the next-to-leading order QCD corrections \cite{MISIAK}
to ~${\rm BR}(b\rightarrow s\gamma)$ ~leaves much less room for new 
physics, which gives an increase in this branching ratio and favours new 
physics which partially cancels the Standard Model contribution to 
the ~$b\rightarrow s\gamma$ ~ amplitude. This partial cancellation
is generic for chargino-stop 
contributions with light sparticles for low ~$\tan\beta$ ~values.

Altogether, recent studies of the electroweak observables in the MSSM 
show that there is 
a particular region in the parameter space with some very light superpartners,
which is consistent with all the available constraints and, as such, of special
interest for the idea of low energy supersymmetry and for experimental searches
at LEP2 and the Tevatron.
\vskip 0.5cm

{\bf 2. SOLUTIONS TO THE RENORMALIZATION GROUP EQUATIONS}
\vskip 0.3cm

We write down first the approximate solutions to the RG equations in the low ~
$\tan\beta$ ~regime (i.e. neglecting the effects of Yukawa couplings other
than the top quark one) for the relevant
parameters. For our purpose, it is useful to give the expression of the
high energy parameters as a function of the low energy ones. They read:
\begin{eqnarray}
\ovm^2(0) 
= {\ovm^2(t)\over 1-y} + y\left(A_t^2(0)-2\hat\xi A_t(0)M_{1/2}\right)
+ {y\hat\eta - y^2\hat\xi^2 - \overline\eta\over 1-y} M_{1/2}^2,
\label{eqn:m2ov0}
\end{eqnarray}
\begin{eqnarray}
m_K^2(0) &=&m_K^2(t) + {c_K\over c}{y\over 1-y}\ovm^2(t)
+\frac{c_K}{c} y \left(A_t^2(0) - 2 \hat\xi A_t(0)M_{1/2}\right)\nn\\
&-& \left[\eta_K + \frac{c_K}{c}{y\over 1-y}\left(\overline\eta
- \hat\eta  + y \hat\xi^2 \right)\right]M_{1/2}^2 + D_K,
\label{eqn:m2k0}
\end{eqnarray}
where
\begin{eqnarray}
\ovm^2(t)\equiv m^2_Q(t) + m^2_U(t) + m^2_{H_2}(t),
\end{eqnarray}
\begin{eqnarray}
D_K = \kappa_K \ovm^2_Y(t)
\left[\left({\alpha_1(0)\over\alpha_1(t)}\right)^{13/33}-1\right],
\label{eqn:dterm}
\end{eqnarray}
\begin{eqnarray}
\ovm^2_Y(t)\equiv -m^2_{H_1}(t) + m^2_{H_2}(t) 
+ \sum_{gen}\left[m^2_E(t)- m^2_L(t) + m^2_Q(t) + m^2_D(t) -2 m^2_U(t)\right]. 
\label{eqn:m2yovt}
\end{eqnarray}
where ~$t\equiv{1\over2\pi}\log{M\over Q}$ ~with ~$M=M_{GUT}$ ~or 
any intermediate scale, and ~$m_K^2$, ~with ~$K = H_i,Q,U,D,L,E$, ~denotes
the soft supersymmetry breaking mass parameters of the Higgs, left-handed 
squark, right-handed up-type squark, right-handed down-type squark, 
left-handed slepton and right-handed slepton, respectively.
Quantities at ~$t=0$ ~are the initial values of the parameters at the 
scale ~$M$. ~Functions ~$\eta_K(t)$, ~$\hat\xi(t)$, ~$\hat\eta(t)$ ~are 
defined in the Appendix
and ~$\overline\eta\equiv\eta_Q+\eta_U+\eta_{H_2}$. ~The evolution of the
trilinear coupling factor ~$A_t$, and of the ~$B$ and ~$\mu$ ~parameters
are given by:  
\begin{eqnarray}
A_t(0) = \frac{A_t(t) + \left(\xi_u - y\hat\xi\right) M_{1/2}}{1-y},
\label{eqn:at0}
\end{eqnarray}
\begin{eqnarray}
B(0)= B(t) + \frac{c_B}{c}\frac{y}{1-y}A_t(t)
+\left[\xi_B +
\frac{c_B}{c}\frac{y}{1-y}\left(\xi_u-{\hat\xi}\right)\right] M_{1/2},
\label{eqn:b0}
\end{eqnarray}
The coefficients ~$c$, ~$c_K$ ~and ~$\kappa_K$ ~read: ~
$c=6$, ~$c_B=3$, ~$c_Q=1$,  ~$c_U=2$, ~$c_{H_2}=3$, ~
$c_L=c_E=c_D=c_{H_1}=0$; ~
$\kappa_{H_1}=-\kappa_{H_2}=\kappa_L = -3/26$, ~$\kappa_E = 3/13$, ~
$\kappa_Q = 1/26$, ~$\kappa_U = -2/13$, ~$\kappa_D = 1/13$. ~
Factors ~$\xi_k(t)$ ~($\xi_B(t)= \xi_e(t)$) ~are defined in the Appendix.
The evolution of the ~$\mu$ ~parameter is given by
\begin{eqnarray}
\mu(0)=\mu(t) E_{\mu}^{-1/2}(t)\left(1-y\right)^{-\left({c_\mu}/{2c}\right)}
\nonumber\\
E_{\mu}(t) = \prod_{i=1,2,3}
\left({\alpha_i(0)\over\alpha_i(t)}\right)^{\left(a^i_{\mu}/b^i\right)},
\label{eqn:mu0}
\end{eqnarray}
where ~$c_{\mu}=3$, ~$a^i_{\mu}=(5/3,3,0)$ ~and ~$b^i=(11, 1, -3)$. ~
The function ~$y\equiv y(t)$ ~is defined as
\begin{eqnarray}
y(t) \equiv {Y_t(t)\over Y_f(t)},
\label{eqn:yt}
\end{eqnarray}
where
\begin{eqnarray}
Y_t(t)= \frac{Y_t(0)E(t)}{1+c Y_t(0) F(t)}
\;\;\;\;\;\;\;\; {\rm and} \;\;\;\;\;\;\;\;\;
Y_f(t)= \frac{E(t)}{c F(t)}.
\label{eqn:yirt}
\end{eqnarray}
Here ~$Y_t = h_t^2/4\pi$, ~where ~$h_t$ ~is the top quark Yukawa coupling 
and the functions ~$E(t)$ ~and ~$F(t)$ ~have the well-known form:
\begin{eqnarray}
E(t) = \prod_{i=1,2,3}
\left(\frac{\alpha_i(0)}{\alpha_i(t)}\right)^{\left(a^i/b^i\right)} 
\;\;\;\;\;\;\;\;\; {\rm and} \;\;\;\;\;\;\;\;\;\;\;
F(t) = \int_0^t E(t^{\prime})dt^{\prime},
\label{eqn:fft}
\end{eqnarray}
with
\begin{eqnarray}
\alpha_i(t) =\frac{\alpha_i(0)}{1+b_i\alpha_i(0) t}.
\label{eqn:alfat}
\end{eqnarray}

The coefficients ~$a^i$ ~are defined in the Appendix. Also 
there, we give the values of ~$\hat\xi(t)$ ~
$\hat\eta(t)$, ~$\xi_i$ ~and ~$\eta_K$ ~for several scales ~$M$ ~and
different values of ~$\alpha_s(M_Z)$. ~

The soft SUSY breaking
parameters are expressed in terms of the physical parameters according to
\begin{eqnarray}
m^2_{H_1}(t) &=& \sin^2\beta M^2_A + {t_{\beta}\over2} M^2_Z - \mu^2 \nn\\ 
m^2_{H_2}(t) &=& \cos^2\beta M^2_A - {t_{\beta}\over2} M^2_Z - \mu^2 \nn\\ 
\mu(t) B(t) &=& \sin\beta\cos\beta M^2_A
\end{eqnarray}
\begin{eqnarray}
m^2_Q(t) &=& M^2_{\tilde t_1}\cos^2\theta_{\tilde t} 
         + M^2_{\tilde t_2}\sin^2\theta_{\tilde t} - m^2_t   
         - {t_{\beta}\over6}\left(M^2_Z - 4M^2_W\right)\nn\\
m^2_U(t) &=& M^2_{\tilde t_1}\sin^2\theta_{\tilde t} 
         + M^2_{\tilde t_2}\cos^2\theta_{\tilde t} - m^2_t   
         +{2\over3}t_{\beta}\left(M^2_Z - M^2_W\right)
\end{eqnarray}
\begin{eqnarray}
A_t(t) = {M^2_{\tilde t_1}-M^2_{\tilde t_2}\over m_t}\sin\theta_{\tilde t},
\cos\theta_{\tilde t} + \mu(t)\cot\beta
\label{eqn:att}
\end{eqnarray}
where ~$M^2_{\tilde t_1}$, ~$M^2_{\tilde t_2}$ ~and ~$\theta_{\tilde t}$ ~
are respectively the heavier and lighter physical top squark masses and their 
mixing angle; ~$t_{\beta}\equiv(\tan^2\beta-1)/(\tan^2\beta+1)$, ~and we 
have ignored the low energy one-loop corrections to the mass parameters 
\cite{PIERCE}, which are 
inessential in determining the qualitative properties of the mass parameters
at the scale ~$M$. ~
One should stress that eqs. (\ref{eqn:m2ov0})-(\ref{eqn:b0}) 
are valid for general, non-universal values of the soft SUSY breaking mass 
parameters at the scale ~$M$ ~and that unification assumptions for the 
gauge couplings and hence for gaugino masses have not been used
($M_{1/2}$ ~is by definition equal to the gluino mass ~$M_3$ ~at the 
scale ~$M$). ~Moreover, the functions ~$Y_f(t)$ ~and ~$y(t)$ ~defined by eqs.
(\ref{eqn:yt})-(\ref{eqn:yirt}) are auxiliary functions defined 
for any scale ~$M$. ~For ~$M=M_{GUT}$ ~a consistent 
perturbative treatment of the theory can only be performed if
\begin{eqnarray}
y_{GUT}\equiv y\left(t={1\over2\pi}\log{M_{GUT}\over M_Z}\right) < 1,
\end{eqnarray}
where ~$y_{GUT}\approx1$ ~defines the quasi-infrared fixed point solution 
\cite{BDM,OP,DYN}. In this limit we obtain the well known 
dependence of ~$m_t$ ~on ~$\tan\beta$ ~which is consistent with 
perturbativity up to the GUT scale \cite{CPW,SEVEN}. 
For scenarios with supersymmetry broken 
at lower scales, ~$M \ll M_{GUT}$, ~the same 
values of ~$m_t$ ~and ~$\tan\beta$ ~(or equivalently of ~
$y_{GUT}$) ~obviously give much lower values for the auxiliary 
function ~$y(t)$, ~where ~$t$ ~is defined, at the scale ~$Q=M_Z$, as~
$t={1\over2\pi}\log{M\over M_Z}$. ~
One should also remember that, in general,
for ~$M < M_{GUT}$, ~new matter multiplets are expected to
contribute to the running of the gauge 
couplings above the scale ~$M$. ~New complete ~$SU(5)$ ~or ~$SO(10)$ ~
multiplets do not destroy the unification of gauge couplings,
but their value at the unification scale becomes larger and,
correspondingly, the perturbativity bound ~$y_{GUT}<1$ ~allows for 
larger values of the top quark Yukawa coupling (smaller ~$\tan\beta$) ~
at the scale ~$M_Z$. ~We shall  use the auxiliary function ~$y(t)$ ~
when presenting our results.
\vskip 0.5cm

{\bf 3. QUALITATIVE FEATURES OF THE SOLUTIONS} 
\vskip 0.3cm

Before proceeding with the analysis,
we recall that our theoretical intuition about the superpartner spectrum
has been to a large extent developed on the top-down approach within
the minimal supergravity model (with universal soft terms) and on
minimal models of low-energy supersymmetry breaking. 
For instance, it is well known that in the minimal supergravity model,
for low ~$\tan\beta$, ~the renormalization group running 
gives\footnote{The coefficients in eq. (\ref{eqn:mz_univ})
               are for ~$\tan\beta=1.6$; they decrease with 
               increasing ~$\tan\beta$. 
               The exact values of the coefficients in eqs.\ 
               (\ref{eqn:mz_univ}),(\ref{eqn:mq_univ}),(\ref{eqn:sumr_phys}) 
               depend only slightly on the values of ~$\alpha_3$, ~ 
               $\sin^2\theta_W$ ~and ~$M_{GUT}$.}:
\begin{eqnarray}
M^2_Z\approx-2\mu^2 + {\cal O}(3) m^2_0 + {\cal O}(12) M^2_{1/2}
\label{eqn:mz_univ}
\end{eqnarray}
(as follows from eqs. (\ref{eqn:m2ov0})-(\ref{eqn:att}) when  
appropriately simplified),
where ~$M_{1/2}$ ~and ~$m_0$ ~are the high energy scale
universal gaugino and scalar mass parameters, respectively. The large 
coefficient in front of ~$M^2_{1/2}$ ~is the well-known source of
fine-tuning for ~$M_{1/2}>M_Z$. ~Therefore, since from (\ref{eqn:mz_univ}) ~
we should have $\mu>M_{1/2}$, ~one naturally expects that
the lightest neutralino and  chargino are both gaugino-like \cite{COPW}. 
For moderate values of the parameters at the GUT scale, 
the RG evolution of the squark masses gives
\begin{eqnarray}
m^2_Q&=&{\cal O}(6) M^2_{1/2} + {\cal O}(0.5) m^2_0 + \ldots
\nonumber\\
m^2_U&=&{\cal O}(4) M^2_{1/2} + \ldots
\label{eqn:mq_univ}
\end{eqnarray}
where the ellipsis stands for terms proportional to ~$m_0^2$, ~
$A_0^2$ ~and ~$A_0 M_{1/2}$ ~($A_0$ ~being the universal trilinear coupling
at the scale ~$M_{GUT}$) ~with coefficients going to zero in the limit ~
$y \to 1$. ~We see that the hierarchy ~$m^2_Q \gg m^2_U$ ~can be generated 
provided ~${\cal O}(1 ~{\rm TeV}) \sim m_0 \gg M_{1/2} \sim {\cal O}(M_Z)$, ~
i.e. consistent with the naturalness criterion.
%
%
%
%
Furthermore, from approximate analytic solutions to the RG evolution, with 
non-universal initial conditions for the scalar soft mass, one can see that
relaxing the universality in the Higgs sector (but still preserving
the ~$SO(10)$ ~relation ~$m^2_Q(0)=m^2_U(0)$) ~is sufficient for generating ~
$m^2_U(t)\simlt0$, ~and also for obtaining solutions with a higgsino-like 
chargino, i.e. for destroying the ~$\mu^2$, ~$M_{1/2}$ ~$(m_0^2)$ ~correlation 
of eq. (\ref{eqn:mz_univ}) \cite{COPW}.

Of course,  {\sl Ans\"atze} for soft terms at the high energy scale, such as 
full or partial universality, select only a small subset of the
low-energy parameter space, even if the latter is assumed to contain
light chargino and stop and to be characterized by the hierarchy ~
$M_{\tilde t_1} \gg M_{\tilde t_2}$. ~Since the sparticle spectrum with ~
$m_{C^+_1}\sim m_{N_1^0} \sim M_{\tilde t_2} \sim {\cal O}(M_Z)$ ~
and ~$M_{\tilde t_1} \gg M_{\tilde t_2}$ ~is important for LEP2, and
at the same time consistent with the precision electroweak data
\cite{MY_MSSM,MY_RB}, it is of interest to study the mapping of such
a spectrum to high energies in a general way and to better understand its 
consistency (or inconsistency) with various simple patterns.

A light-right handed stop, with mass of the order of ~$M_Z$, ~is also 
consistent with scenarios in which the observed baryon number of the Universe
is generated at the electroweak phase transition. It is well known
that the baryon number asymmetry generated at energy scales of the order of
the grand unification scale may be efficiently erased by unsuppressed
anomalous processes at temperatures far above the electroweak phase
transition. If this is the case, the baryon number must have been generated
by non-equilibrium processes at a sufficiently strongly
first-order electroweak phase transition.
If all sparticles are heavy, the effective low energy theory is equivalent
to the Standard Model, in which the phase transition is too
weakly first order to allow for this possibility. The presence of a light
stop, instead, has a major impact on the structure of the finite-temperature
Higgs effective potential and, indeed, it has been shown that it is
sufficient to increase the strength of the first-order phase transition,
opening a window for electroweak baryogenesis \cite{QUIROS}. Moreover, light 
stops and light charginos may provide the new $CP$-violation phases necessary
to efficiently produce the observed baryon asymmetry \cite{HN}. For this 
scenario to work, ~$\tan\beta$ ~must take values close to ~1, ~and the Higgs 
mass must be within the reach of the LEP2 collider.  

We shall now discuss some general features of the behaviour of the soft 
supersymmetry breaking parameters, which may be extracted from 
eqs. (\ref{eqn:m2ov0})-(\ref{eqn:mu0}). 
We consider first the case ~$M=M_{GUT}$ ~and
the limit ~$y_{GUT}\rightarrow1$ ~as a useful reference frame. One
should be aware, however, that already for ~$y_{GUT}\simeq 0.8$--$0.9$ ~large
qualitative departures from the results associated with
this limit are possible.
In the limit ~$y_{GUT}\rightarrow1$, ~from eqs. (\ref{eqn:m2ov0}) and 
(\ref{eqn:at0}), we see that if the high energy parameters are of the same
order of magnitude as the low energy ones, the following relations must
be fulfilled \cite{COPW}:
\begin{eqnarray}
\Delta_{m^2}&\equiv&
\ovm^2(t)-\left(\overline\eta -y\hat\eta
+ y^2 \hat{\xi}^2\right)M^2_{1/2}\rightarrow0\nn\\
\Delta_A&\equiv&A_t(t)-\left(y\hat\xi -\xi_u\right) M_{1/2}\rightarrow0 ,
\label{eqn:sumrule}
\end{eqnarray}
irrespectively of their initial values (IR fixed point). This is a 
well-known prediction, which remains valid, for ~$y_{GUT}\neq1$, ~in  
the gaugino-dominated supersymmetry breaking scenario in the minimal 
supergravity model.  

Let us now suppose that the relations (\ref{eqn:sumrule}) are strongly violated
by experiment, namely that the scalar masses (or the soft supersymmetry
breaking parameter ~$|A_t|$) ~are very different from those predicted by
(\ref{eqn:sumrule}), for values of ~$m_t$ ~and ~$\tan\beta$ ~corresponding
to ~$y_{GUT}$ ~close to ~1. ~Clearly, this means that
\begin{eqnarray}
A_t(0)&\sim & {\cal O}\left({\Delta_A\over1-y}\right)\nn \\
\overline m^2(0) 
&\sim &{\cal O}\left({\Delta_{m^2}\over1-y}\right)+ {\cal O}(A_t^2(0))  
\end{eqnarray}
i.e. supersymmetry breaking must be driven by very large initial scalar masses
and the magnitude of the effect depends on the departure from the 
fixed point relations for ~$A_t$ ~and ~$\ovm^2$, ~eq.
(\ref{eqn:sumrule}), ~and the proximity to the top-quark mass infrared 
fixed point solution measured by ~$1-y_{GUT}$. ~Moreover, for not very 
small values of ~$|\Delta_A|$ ~and/or ~$|\Delta_{m^2}|$, ~it follows 
from eq. (\ref{eqn:m2k0}) that in the limit ~$y_{GUT}\rightarrow1$, ~ 
the initial values are correlated in such a way that
\begin{eqnarray}
m^2_Q(0)~ :~ m^2_U(0)~ :~m^2_{H_2}(0) ~\simeq~1 ~:~ 2~ :~ 3, 
\label{eqn:fprel}
\end{eqnarray}
independently of the actual values of those masses at the scale ~$M_Z$. 
In other words, the values at ~$M_Z$ ~are obtained (via RGEs) by a very high 
degree of fine-tuning between the initial values ~$m^2_K(0)$. ~
We conclude that eqs. (\ref{eqn:sumrule}) are necessary (but of course 
not sufficient) conditions for large departures from the prediction of eq.
(\ref{eqn:fprel}) for the soft scalar masses in the limit ~$y\rightarrow1$. ~
In particular, they are necessary conditions to render the spectrum consistent
with fully (or partially) universal initial values of the third-generation 
squark and Higgs boson soft masses. In Figs. 1 and 2 we plot the  ~
$\Delta_{m^2}/\sqrt{|\Delta_{m^2}|}$ ~as a function of ~$r\equiv M_2/|\mu|$ ~
and the contours of ~$\Delta_A$ ~in the plane ~$(r, ~\theta_{\tilde t})$ 
respectively ~for ~$\tan\beta=1.6$, ~generic values of the low energy 
parameters ~$M_{\tilde t_1}$, ~$M_A$ ~and with ~
$M_{\tilde t_2} \sim m_{C^+_1} < 100$ GeV. ~We note that the conditions
(\ref{eqn:sumrule}) are simultaneously satisfied only in very restricted
regions of the low energy parameter space, which are fixed by the following
equations:
\begin{eqnarray}
\Delta_{m^2}&\equiv&\cos^2\beta M^2_A + M^2_{\tilde t_2} + M^2_{\tilde t_1} 
- \cos2\beta M^2_Z - \mu^2 - {\cal O}(6) M^2_{1/2} = 0\nonumber\\
\Delta_A&\equiv&{M^2_{\tilde t_1} - M^2_{\tilde t_2}\over 2m_t}
\sin2\theta_{\tilde t} + \mu\cot\beta + {\cal O}(2) M_{1/2} = 0
\label{eqn:sumr_phys}
\end{eqnarray}
The existence of two regions in ~$r\equiv M_2/|\mu|$, ~where eqs.
(\ref{eqn:sumrule}) can be satisfied and the difference between positive and 
negative values of ~$\mu$ ~is explained by eqs. (\ref{eqn:sumr_phys}) and
Figs. 2. ~In Figs. 2 we also plot the contours of ~$\delta R_b$ ~to be
discussed later.

When eqs. (\ref{eqn:sumrule}) are violated, there 
are essentially two ways of departing from the prediction of eq.
(\ref{eqn:fprel}). One is to increase the value of ~$\tan\beta$ ~
(i.e. to decrease the top-quark Yukawa coupling for fixed ~$m_t$). ~
The other way is to lower the scale at which supersymmetry breaking is 
transferred to the observable sector, since for ~$M\ll M_{GUT}$ ~the
soft supersymmetry breaking parameters do not feel the strong rise
of the top quark Yukawa coupling at scales close to ~$M_{GUT}$. ~In both
cases, ~$y$ ~takes values smaller than ~1 ~and we can depart from 
(\ref{eqn:fprel}) even when (\ref{eqn:sumrule}) are not satisfied.

For any value of ~$y$, ~
one has the following set of equations, which relate the low-energy
supersymmetry-breaking parameters to their high energy values:
\begin{eqnarray}
&&3 m_Q^2(0) - m_{H_2}^2(0)=
3 m_Q^2(t) - m_{H_2}^2(t) - \left(3 \eta_Q - \eta_{H_2}\right) M_{1/2}^2
\nonumber\\
&&2 m_Q^2(0) - m_{U}^2(0) =
2 m_Q^2(t) - m_{U}^2 (t)- \left(2 \eta_Q - \eta_U \right) M_{1/2}^2 
+ (2 D_Q(t) - D_U(t)),\nonumber\\
\label{eqn:line}
\end{eqnarray}
where ~$D_Q$ ~and ~$D_U$ ~are the {\it D}-term contributions to the evolution
of the mass parameters ~$m_Q^2$ ~and ~$m_U^2$ (eq. (\ref{eqn:dterm})). ~
The above relations, eq. (\ref{eqn:line}), 
depend on the scale ~$M$ ~through the coefficients ~$\eta_K$. ~ 
For a definite scale ~$M$ ~they define a line in the ~
($m_Q^2(0)$, $m_U^2(0)$, $m_{H_2}^2(0)$) ~space, which is parallel
to the line ~$m_Q^2(0)=w$, ~$m_U^2(0)=2w$, ~$m_{H_2}^2(0)=3w$ ~($w$ ~being the 
parameter of the line). Thus, for values at the scale ~$M$ ~much larger than 
the low energy masses of the scalar mass parameters, the points
on this line fulfil the relation ~$m_{H_2}^2:m_U^2:m_Q^2 \to 3:2:1$. ~
The high energy solution for the mass parameters may be obtained
by the intersection of the line, eq. (\ref{eqn:line}), 
with the plane defined by the
relation (\ref{eqn:m2ov0}). It is easy to find a geometrical
interpretation for the behaviour of the solutions close to the 
top-quark mass fixed point 
solution. If  the fixed point relations, eq. (\ref{eqn:sumrule}), are not 
fulfilled, the distance of the plane 
defined by eq. (\ref{eqn:m2ov0}) to the origin increases 
as ~$y$ ~approaches ~1 ~and, hence, the high energy solutions lie
in the asymptotic regime of the 
line defined by eq. (\ref{eqn:line}). Hence, the soft scalar
masses at the scale ~$M$ ~fulfil the ~$3:2:1$ ~relation discussed 
above. On the contrary, if the fixed point relations, of
eqs. (\ref{eqn:sumrule}) are approximately fulfilled, 
when ~$y\rightarrow1$ ~any point in the line defined by 
eq. (\ref{eqn:line}) will lead to a solution consistent with
the low energy parameters. This property reflects the presence of
a fixed point solution for ~$\ovm^2$, ~that is the low-energy solution
becomes independent of ~$\ovm^2(0)$. ~Moreover, when the relations defined in  
eqs. (\ref{eqn:sumrule}) are not fulfilled, but ~$y$ ~is sufficiently smaller 
than ~1, ~for the quantity ~$\ovm^2(0)$ ~to remain small, the plane
given by eq. (\ref{eqn:m2ov0}) crosses the line, eqs. (\ref{eqn:line}), 
at small values of ~$m^2_Q(0)$, ~$m^2_U(0)$ ~and ~
$m^2_{H_2}(0)$. ~The relation between these 
initial values then depends on the exact location of the line defined by eqs.
(\ref{eqn:line}), which in turn depends on the low energy values of ~
$m^2_Q$, ~$m^2_U$, ~$m^2_{H_2}$ ~and ~$M_{1/2}$. ~The regions corresponding
to universal (or partially universal) pattern for the high-energy scale 
soft terms can then be found.
\vskip 0.5cm

{\bf 4. MAPPING TO THE GRAND UNIFICATION SCALE} 
\vskip 0.3cm

Guided by the qualitative considerations presented in section 3,
we shall present the bottom-up
mapping of the low energy parameter space characterized as  follows:
\begin{eqnarray}
50 \; {\rm GeV} \simlt & M_{\widetilde{t}_2} & \simlt 100 ~{\rm GeV}
\nonumber\\
80 \; {\rm GeV} \simlt & m_{C^+}  & \simlt 100 {\rm ~GeV}
\nonumber\\
0.1 < & r\equiv M_2/|\mu| & < 10 
\end{eqnarray}
for both signs of $\mu$. 
To reduce the parameter space, in our numerical analysis we make the 
assumption that
\begin{eqnarray}
{M_1\over\alpha_1}={M_2\over\alpha_2}={M_3\over\alpha_3}
\label{eqn:unass}
\end{eqnarray}
at any scale. For ~$m_t=175$ GeV, ~$\alpha_s(M_Z)=0.118$ ~and ~
$M=M_{GUT}=2\times10^{16}$ GeV ~we consider 
\begin{eqnarray}
0.98 \geq & y_{GUT} & \geq 0.80,
\end{eqnarray}
corresponding to ~$\tan\beta$ ~in the range ~1.6--2.8 ~ in the approximation 
\begin{equation}
Y_t(M_Z)={1 \over 2}{\alpha_2(M_Z) \over \cos^2\theta_W}
{m_t^2 \over M_Z^2} \left( 1 + \cot^2\beta \right).
\end{equation}
As an example of a low scale we take ~$M= 10^7$ GeV, ~fow which 
values of ~$y(t)=0.75$ ~and ~0.50 ~are 
considered, which correspond to ~$\tan\beta = 1.25$ ~and ~3.7, ~respectively.
The second, rather large value of ~$\tan\beta$, ~has been chosen in order
to stress that the solutions are rather insensitive to the
value of ~$y$, ~provided ~$y\ll 1$, ~and their applicability is limited only 
by the validity of our approximate solutions to RGEs. Our choice for
the scale $M$ will be justified in section 5.

We scan over the remaining relevant parameters ~$M_A$, ~
$M_{\tilde t_1} < 1$ TeV, ~and the range of ~$|\theta_{\tilde t}|< 60^o$. ~
The accepted low-energy parameter space is then defined as a subspace
in which ~$M_h>60$ GeV, ~$0.98\times10^{-4} < BR(b\rightarrow s\gamma) 
<3.66\times10^{-4}$; ~we also 
require that ~$\chi^2\leq\chi^2_{min ~SM} + 2$, ~
where the ~$\chi^2$ ~fit includes the electroweak observables, which do not
involve heavy flavours. In particular, the stop mixing angle ~
$\theta_{\tilde t}$ ~is directly related to ~$M_h$ ~\cite{HEM}, 
while in general its sign becomes related to the sign of ~$\mu$ ~through the
rate of ~$b\rightarrow s\gamma$
\footnote{In our computations we simulated the full next-to-leading results 
          for ~$BR(b\rightarrow s\gamma)$ ~(which reduce the theoretical 
          error associated with the choice of the final RG evolution scale) 
          \cite{MISIAK} by narrowing appropriately the variation of the 
          final scale $Q\sim {\cal O}(m_b)$, 
          so that it reproduces higher-order results.}. ~
The range of ~$y_{GUT}$ ~allows a discussion of
the effect of the infrared fixed point and departures from it;
the ratio ~$M_2/\mu$ ~allows to study different compositions 
of the lightest chargino.
Moreover, heavy ~${\tilde t}_L$ ~is necessary to avoid unwanted
positive contributions to ~$\Delta\rho$ ~and to keep
the Higgs mass above the experimentally excluded range. 
Since the pattern of soft supersymmetry breaking parameters does not
depend strongly on the exact value of the light chargino and stop masses,
we present our results for ~$m_{C^+} = 90$ GeV, ~$M_{\tilde t_2} = 60$ GeV, ~
$y_{GUT} = 0.98$ and  $0.80$ ~and for both signs of ~$\mu$. ~ 

The regions in the low-energy parameter space allowed by the experimental 
constraints listed above are shown in Figs. 3 and 4 ~as projections on 
the ~$(M_{\tilde t_1}, ~\theta_{\tilde t})$ ~and ~$(r, ~M_A)$ ~planes, for ~
$y=0.98$ and 0.80. ~Also, in Figs. 5, ~we show similar regions for ~$y=0.75$ ~
and ~$M=10^7$ GeV. ~
We observe the qualitatively expected behaviour of the bounds. In particular,
the limits in the ~$(M_{\tilde t_1}, ~\theta_{\tilde t})$ ~plane show the 
expected strong bound on ~$M_{\tilde t_1}$ ~correlated with the value of ~
$\theta_{\tilde t}$. ~It follows both from the limits on ~$M_h$ ~(under the
assumption of a small $M_{\tilde t_2}$) ~and from the constraints on ~
$\Delta\rho$. ~Moreover, the vanishing left--right mixing
angle is excluded for ~$\tan\beta<2$ ~by the same constraints. Lower bounds 
on ~$M_A$ ~as a function of ~$r$ ~show an increase for gaugino-like
charginos ~($r<1$), ~which follows from 
the ~$b\rightarrow s\gamma$ ~constraint. In the region of 
parameter space allowed by the experimental constraints, 
we are generally far away from the fixed point 
relations,  eqs. (\ref{eqn:sumrule}). For ~$M = M_{GUT}$, ~for which ~
$y\simgt 0.80$,  ~the global pattern
of soft supersymmetry breaking parameters is hence governed
by the strong deviation from their infrared fixed point solution
and the largeness of the top-quark Yukawa coupling. 
The dependence of the ~$m^2_Q$, ~$m^2_U$, ~$m^2_{H_2}$ 
and ~$m^2_{H_1}$ ~renormalization group evolution 
on ~$m^2_L$, ~$m^2_E$, ~$m^2_D$ ~and ~$A_b$, which enters only through the ~
{\it D}-term ~contributions, eq. (\ref{eqn:dterm}), is taken into account
by considering five different 
values ~$\ovm^2_Y(t) =0,\pm5\times10^5,\pm10^6$ $({\rm GeV})^2$.~

Figures 6 and 7 show the results for the soft supersymmetry
breaking parameters at the scale ~$M = M_{GUT}$, ~obtained by mapping 
the whole low energy region shown in Figs. 3 and 4. ~We plot
\begin{equation}
m_K\equiv {m^2_K(0)\over\sqrt{|m^2_K(0)|}},\;\;\;\;\;\;\;A_t\equiv A_t(0),
\;\;\;\;\;\;\;B\equiv B(0),
\end{equation}
for two different values
of the top quark Yukawa coupling, ~$y\simeq 0.98$ ~and ~
$0.8$. ~Since the chargino mass is fixed, for any
value of ~$r = M_{2}/|\mu|$, ~the values of ~$M_2$ ~and ~$\mu$ ~are
uniquely determined (for any given sign of ~$\mu$). ~We have checked that
the qualitative behaviour of the solutions
do not strongly depend on the sign of ~$\mu$.

Figure 6a shows that, for ~$y\simeq 1$, ~the characteristic 
third-generation squark scalar
masses at the scale ~$M_{GUT}$ ~must be, in general, much larger than the 
common gaugino masses. The stronger the deviation from the low-energy
fixed-point relations, eqs. (\ref{eqn:sumrule}), the larger the value
of ~$m_Q/M_{1/2}$. ~The strong correlations between the different
scalar mass parameters at ~$M_{GUT}$, ~eq. (\ref{eqn:fprel}), 
ensure the strong
cancellations necessary to obtain low energy squark masses of
order ~$M_{1/2}$. ~For values of ~$y$ not very close to ~$1$, ~the resulting
values of ~$m_Q/M_{1/2}$ ~become smaller and the correlations are weaker. 
This behaviour is clearly illustrated in Fig. 7a ~for ~$y=0.80$. ~

In Fig. 6b we show the behaviour of ~$m_U/m_Q$ ~
for different values of ~$M_{1/2}$ ~selected in our scan.
A clear concentration of solutions around ~$m_U/m_Q = \sqrt{2}$ ~
appears. This concentration dilutes when
we depart from the infrared fixed point solution for the top quark
mass, as is clearly shown in Fig. 7b. 
Due to the relation of eq. (\ref{eqn:line}), for a given stop spectrum,  the
values of ~$m_Q/m_U$ ~are mainly governed by the size of ~$M_{1/2}$. ~
For low values of ~$M_{1/2}$, ~the maximum value
of ~$m_U/m_Q$ ~is given by $\sqrt{2}$, while for all solutions with large
values of ~$M_{1/2} = {\cal O}$(1 TeV) one has ~$m_U/m_Q > \sqrt2$. ~This is
clearly seen in Figs 6b and 7b, where all solutions converge to ~
$m_Q/m_U = \sqrt{2}$ ~for large values of ~$m_Q/M_{1/2}$. ~

In Figs 6c and 7c ~we show the behaviour of the soft supersymmetry breaking
Higgs mass parameters. Similarly to the relation between ~$m_U$ ~and ~$m_Q$, ~
it follows from eq. (\ref{eqn:line}) that the hierarchy between ~
$m_{H_2}$ ~and ~$m_Q$ ~is governed by the size of the gaugino masses.
Values of ~$m_{H_2}/m_Q\geq\sqrt3$ ~may only appear for large
values of the gaugino masses, while low values always 
lead to ~$m_{H_2}/m_Q\leq\sqrt3$. ~For ~$y\simeq 1$, ~a strong concentration
of solutions around the boundary value is observed. Solutions with
negative values of ~$m^2_{H_1}$, ~associated with large values of ~$\mu$, ~
are present in the data. For values of the lightest stop mass  of order 
or below ~$M_Z$, ~a non-negligible stop mixing is in general necessary. 
For moderate values of ~$\mu$ ~
and ~$M_{1/2}$, ~values of ~$A_t(t)$ ~appear, which violate  
relations (\ref{eqn:sumrule}). Therefore,
from eqs. (\ref{eqn:at0}), it follows that  close to the fixed
point for the top quark mass, ~$A_t$ ~becomes large.  From 
eq. (\ref{eqn:m2k0}), it follows that, in this case, ~$A_t$ ~
dominates the renormalization group evolution of the scalar mass, implying
the correlation ~$A_t/m_Q\simeq\sqrt6$, ~which is clearly
observable in Fig. 6e. For lower values of ~$y$, ~ 
as displayed in Fig. 7e, ~$A_t$ ~ceases to
dominate the evolution of the scalar mass parameters. 

The behaviour of the solutions shown in Figs. 6 and 7 is just a reflection
of the properties discussed above and shows the global tendency in the
mapping of the low energy region selected by our criteria. The last important 
point we would like to discuss concerns the properties of the subregion,
in the low-energy parameter space, that is consistent with simple patterns
of the high-energy soft SUSY breaking parameters, such as universality
or partial universality of scalar masses. It is clear from our earlier
discussion that the closer~is ~$y$  to ~$1$ ~the more restrictive such an
{\sl Ansatz} is. 

In Figs 3 and 4 ~we show the regions,
in the ~$(\theta_{\tilde t}, M_{\tilde t_1})$ ~and ~$(M_A, r)$ ~planes,
which are consistent with the experimental constraints and can be obtained ~
a) with fully universal\footnote{We take into 
          account possible high-energy threshold corrections to
          initial conditions by allowing deviations of the order of 10\% from 
          the assumed equality of the relevant masses squared.}
scalar masses at ~$M_{GUT}$, ~b) with ~$m^2_Q=m^2_U=m^2_{H_2}$, ~
c) with ~$m^2_Q=m^2_U$ ~($SO(10)$-type boundary conditions).
As expected from our qualitative analysis in section 3 and from Figs.
1 and 2, ~such regions indeed exist, even for ~$y\rightarrow1$. ~
Cases a) and b) can only be realized  with gaugino-like charginos and with 
large ~$A_t$. ~In the case c) we get both gaugino- and higgsino-like charginos
(in agreement with the results displayed in  Figs. 1 and 2)  
and the Higgs soft SUSY breaking masses tend 
to be ordered in such a way that ~$m^2_{H_2} > m^2_Q > m^2_{H_1}$. ~
Comparison with contours in Figs. 2  shows that solutions with
higgsino-like charginos fall into the region that gives ~$R_b$ ~larger
than in the SM by ~$\delta R_b\sim0.0010$. ~Larger positive deviations from
the Standard Model predictions for ~$R_b$ ~demand both
larger values of ~$m_Q/M_{1/2}$ ~and soft supersymmetry breaking
mass parameters with a hierarchy of masses close to the one given by
eq. (\ref{eqn:fprel}).
Similar conclusions, although less sharp, hold also for ~$y=0.80$. ~
We conclude that in the low-energy parameter space characterized by 
light chargino and stop there exists a subspace corresponding to 
the simple patterns of initial conditions discussed above which is
consistent with the available experimental data. 
The hierarchy ~$m_0\gg M_{1/2}$ ~must be, of course, present if 
the scalar masses at ~$M_{GUT}$ ~are to satisfy ~
$m^2_Q\approx m^2_U\approx m^2_{H_2}$ ~or ~
$m^2_Q\approx m^2_U\approx m^2_{H_2}\approx m^2_{H_1}$ ~
In the case of ~$SO(10)$-type boundary conditions, the hierarchy ~
$m_0\gg M_{1/2}$ ~is also generically present although there can  
be also solutions with ~$m_0\simlt M_{1/2}$.
\vskip 0.5cm

{\bf 5. GAUGE-MEDIATED SCENARIOS OF SUSY BREAKING}
\vskip 0.3cm

Models in which supersymmetry breaking is transmitted to the
observable sector through ordinary ~$SU(3)\times SU(2)\times U(1)$ ~
gauge interactions of the so-called messenger fields at scales ~
$M\ll M_{GUT}$ ~have recently attracted a lot of interest 
\cite{DINE,KANE1,KOWI,ALEX,DIMGIU}. 
In general, these gauge-mediated models of SUSY breaking are characterized
by two scales: the scale ~$M$, ~which is of the order of the average messenger
mass and the scale ~$\sqrt{F}$ ~of supersymmetry breaking.
Since the physical messenger masses squared are  given by
\begin{equation} 
m_{\rm mess.}^2 \simeq M^2 \pm F,
\end{equation} 
one must have ~$F < M^2$. ~In the minimal models a
lower bound on the scale ~$M$, ~$M\simgt20$ TeV, ~is set by the requirement ~
$M_{\tilde\nu}$, $M_{\tilde e_R} > M_Z/2$ ~\cite{DIMGIU}.
In those models the ordinary ``LSP'' always decays into a gravitino (and
an ordinary particle) whose mass is related to the scale ~$\sqrt{F}$ ~by
\begin{eqnarray}
m_{\tilde G}\simeq 4 {\rm ~eV} \times 
\left({\sqrt{F}\over100 ~{\rm TeV}}\right)^2.
\end{eqnarray}
Since the decay length of the lightest neutral
gaugino into photon and gravitino is approximately given by
\begin{equation}
{\it l} \simeq 2 \times 10^{-3} \times 
\left({\sqrt{F}\over75 ~{\rm TeV}}\right)^4,
\end{equation}
it follows that, if ~$\sqrt{F}\simlt10^6$ GeV, ~apart from
missing energy, photons become a relevant signature for supersymmetry
search at LEP and Tevatron colliders, because the lightest neutralino
decays before leaving the detector.
Such final photon states may be efficiently searched for at the
Tevatron collider, and the absence of any spectacular signature
in the existing ~$\sim100$ pb$^{-1}$ ~of data
strongly disfavours chargino and stop masses below ~100 GeV ~ \cite{KANE1}.
Hence, the chargino and stop would be outside the reach of the LEP2 collider. 
Since our motivation is based on LEP2 physics, we consider here the 
scales ~$M$ ~and ~$\sqrt{F}$ ~such that the neutralino always decays only 
after leaving the  detector. This leads to values ~$\sqrt{F}\simgt10^6$ GeV ~
and ~$M\simgt10^7$ GeV. ~We have also analysed what happens 
for ~$M\approx10^5$ GeV ~(usually considered
in the context of gauge-mediated models \cite{DINE,KOWI,KANE1}), 
and values of the stop and chargino masses consistent
with the Tevatron bounds. We shall comment on this
possibility at the end of this section.

Figures 8 and 9 ~show the solutions obtained for ~$M = 10^7$ GeV ~and two 
different values of the auxiliary function ~$y = 0.75$ ~and ~$y = 0.50$, ~
which, for ~$m_t=175$ GeV ~and ~$\alpha_s(M_Z)=0.118$, ~correspond to the 
low (1.2) and moderate (5.9) ~$\tan\beta$ ~regimes, respectively.
The variation with ~$\tan\beta$ ~of the soft
supersymmetry-breaking parameters is weaker than for ~
$M=M_{GUT}$, ~reflecting the lower sensitivity to the GUT scale
fixed-point solution. ~The dependence on the actual value of ~
$y_{GUT}$ ~is  weaker, since the auxiliary function ~$y(t)$ ~always takes
values smaller than ~1. ~For instance, for ~$y_{GUT}=1$ ~
($M_{GUT}=2\times10^{16}$ GeV) ~and ~$M=10^7$ TeV, we get ~$y(t)\sim 0.70$. ~
As we explained before, since additional matter multiplets must be present
at scales above the scale $M$, ~
a value of $y(t) \sim 0.75$ does not imply a breakdown of the perturbative 
consistency of the theory at scales below the grand unification scale. 
In this case the violation of the sum rules, eq. (\ref{eqn:sumrule}), 
is not associated with an enhancement of the values of the parameters
at the scale ~$M$. ~From the analysis of our equations, for these low
values of ~$M$ ~and ~$y(t)$, ~it follows that 
the values of the soft terms at the scale ~$M$~
tend to reflect the pattern observed at the ~$M_Z$ ~scale. Still,
the larger ~$y(t)$ ~the closer we are to the effects discussed earlier
and the larger the modification of the pattern at ~$M_Z$ ~with respect to
the initial conditions. Indeed, as can be seen from the comparison of
Figs. 7 and 8, the qualitative behaviour of the solutions for ~$y = 0.75$ ~
is very similar to the one obtained in the gravity mediated supersymmetry
breaking scenario for ~$y=0.80$.

Due to the smaller renormalization group running effects
and the fact that ~$\eta_Q\simeq 1.6$, ~the ratio ~
$m_Q/M_{1/2}$ ~reflects the chosen values of
the left-handed stop parameter and the gluino mass at 
low energies, increasing for lower values of ~$M_{1/2}$, ~as is clearly
shown in Figs. 8a and 9a.
More interesting is the behaviour of ~$m_U/m_Q$ ~displayed in
Figs. 8b and 9b. The concentration of solutions around ~
$m_U/m_Q \simeq \sqrt{2}$ ~is still visible at low values of 
$\tan\beta$, but it disappears for  low values of
$y$. Indeed, for $y = 0.50$
the ratio ~$m_U/m_Q$ ~tends, in general, to be lower than ~
1. ~However, values of ~$m_U/m_Q$ ~of order ~1 ~can also appear as possible 
solutions in this scenario. Hence, a light stop is not necessarily
in conflict with models of supersymmetry breaking in which ~
$m_U/m_Q\simeq 1$ ~at the scale ~$M\sim10^7$ GeV.
This is true for both values of ~$y$, ~
although it is easier to accommodate ~$m_U/m_Q$ ~of order ~1 ~for
low values of ~$\tan\beta$ ~(larger values of ~$y$). ~
For $y = 0.5$, however, ~$m_U/m_Q\sim1$ ~implies large values of ~$A_t$ ~
and/or large values of ~$M_A$. ~

For ~$M=10^7$ GeV, ~instead of going into further detail of the mapping
of the full allowed low-energy region, it is more interesting to check
where there are points that correspond to initial values of the soft
parameters, which are generic for the gauge-mediated scenarios of SUSY
breaking. 

We impose the most universal conditions in the following way. 
In the more restrictive 
version\footnote{For definiteness, we consider here the minimal messenger 
                 model and assume that ~$F/M^2\ll 1$ ~so that the approximate
                 formulae (\ref{eqn:m12_gmed}) and (\ref{eqn:mq_gmed}) are
                 valid.}, 
{}from
\begin{eqnarray}
M_{1/2} = {\alpha_3(0)\over4\pi}\Lambda
\label{eqn:m12_gmed}
\end{eqnarray}
we extract ~$\Lambda=F/M$ ~and then ~$m^2_Q(0)$ ~and ~$m_U^2(0)$ ~
are required to be no more than 10\% off from the values
\begin{eqnarray}
m^2_{Q ~g.m.}&=&2\Lambda^2\left[{4\over3}\left({\alpha_3(0)\over4\pi}\right)^2
                             + {3\over4}\left({\alpha_2(0)\over4\pi}\right)^2
                      + {1\over60}\left({\alpha_1(0)\over4\pi}\right)^2\right]
\nonumber\\
m^2_{U ~g.m.}&=&2\Lambda^2\left[{4\over3}\left({\alpha_3(0)\over4\pi}\right)^2
                    + {4\over15}\left({\alpha_1(0)\over4\pi}\right)^2\right]
\label{eqn:mq_gmed}
\end{eqnarray}
predicted in such models \cite{DINE}
\footnote{For an analysis of the renormalization group evolution of the mass 
          parameters in gauge-mediated supersymmetry breaking models, see, 
          for example, Ref. \cite{KOWI}.}. 
In the less restrictive version, ~$m_Q^2(0)$ ~and ~
$m^2_U(0)$ ~must only satisfy (within 10\%) the relation
following from eq. (\ref{eqn:mq_gmed}) and we do not impose condition
(\ref{eqn:m12_gmed}).
We do not constrain ~$m^2_{H_{1,2}}(0)$ ~or ~$A_t$. ~
In gauge-mediated models the prediction ~$A_t\sim0$ ~is also quite 
general, but we prefer to see the regions defined by the conditions
specified above and to check the corresponding values of ~$m^2_{H_{1,2}}(0)$ ~
and ~$A_t$. ~

In Figs. 5a,b we show the contours consistent with experimental 
constraints and mark the regions defined by the above conditions (more
and less restrictive cases are shown by dark stars and white rhombs,
respectively), whereas in Fig. 10 ~we show the corresponding 
ratios ~$h_{1,2}$ ~of 
the actual values of ~$m^2_{H_{1,2}}(0)$ ~obtained from the low-energy
mapping to the values  ~$m^2_{H_{1,2} ~g.m.}$ ~
predicted in the minimal models of gauge-mediated symmetry breaking
(with ~$\Lambda$ ~determined from eq. (\ref{eqn:m12_gmed})
or (\ref{eqn:mq_gmed}) in the more and less restrictive cases, respectively): 
\begin{eqnarray}
m^2_{H_{1,2} ~g.m.}=
2\Lambda^2\left[{3\over4}\left({\alpha_2(0)\over4\pi}\right)^2
          + {3\over20}\left({\alpha_1(0)\over4\pi}\right)^2\right].
\end{eqnarray}
Also, the obtained values of ~$A_t$ ~and ~$B$ ~are displayed. 
{}From Fig. 10, it is clear that a light stop is always associated
with values of ~$m_{H_2}^2(0)$ ~very large with respect to the ones that 
would be obtained in the minimal models. However, because of the 
uncertainty in the values of these parameters associated with the
physics involved in the solution to the ~$\mu$ ~problem in the 
gauge-mediated supersymmetry-breaking models \cite{ALEX}, such 
large values of ~$m^2_{H_2 ~g.m.}$ ~are not necessarily inconsistent  
in this context. The value of ~$A_t$ ~must be, in general, 
large, a reflection of the small hierarchy of scales and the
large values of the stop  mixing necessary to generate a light stop.
Smaller values of ~$A_t$ ~can only be obtained by increasing the value of ~
$m^2_{H_2}$ ~and, hence, the mass of the $CP$-odd Higgs boson. 
In fact, for $m_U(0) \simeq m_Q(0)$, $A_t \simeq 0$, we obtain
\begin{eqnarray}
M_A^2&=&{M_Z^2\over2}\left(\tan^2\beta -1\right) +
\left[\mu^2+m_Q^2(t)\;{6-7y\over y}+ m_U^2(t) \; {5y-6\over y}\right.
\nonumber\\
&+&\left. M_{1/2}^2\left(\eta_Q\; {7y-6\over y}+\eta_U \; \frac{6 - 5y}{y} 
+\eta_{H_2} - \hat{\eta} + 
y \hat{\xi}^2 \right) \right] \left(\tan^2\beta+1\right).
\label{eqn:a_mass}
\end{eqnarray}
It is easy to show that the ~$M_{1/2}^2$ ~coefficient is positive for any
value of ~$M$ ~and ~$y$. ~It follows from the above expression that,
for ~$y<0.8$, ~values of ~$m_U^2 \ll m_Q^2$ ~can only be obtained 
for large values of ~$M_A$, ~particularly for large values of ~$\tan\beta$. ~
Indeed, in the more restrictive case, for ~$y=0.75$ (0.50), ~i.e. for ~
$\tan\beta$ ~small (intermediate), solutions with ~$A_t\sim0$ ~
are possible only for very large values of ~$M_A$, ~$M_A\simgt1 ~(5)$ TeV, ~ 
and ~$h_{1,2}\sim {\cal O}(15)$. ~
In the less restrictive case, for ~$y=0.75$, ~solutions with ~$A_t\sim0$ ~
are possible already with ~$M_A\simgt800$ GeV (they require, however, ~
$h_{1,2}\sim {\cal O}(20)$). ~For ~$y=0.50$, ~even in the less restrictive
case, they require extremely large values of ~$M_A$ ~($M_A\simgt 3$ TeV) and ~
$h_{1,2}\simgt {\cal O}(50)$. ~

Let us end this section by mentioning what happens for ~$M\approx10^5$ GeV ~
and values of the chargino and stop masses slightly above the present 
Tevatron bounds. Notice that, although this range of masses is
outside the reach of LEP2, it is still consistent with the requirements
of electroweak baryogenesis in the MSSM. In this case, for values of ~
$\tan\beta$ ~similar to the ones chosen for ~$M=10^7$ GeV, ~one obtains the 
values ~$y=0.55$ ~and ~$y=0.35$, ~
respectively, while the coefficients associated with the dependence
of the scalar masses on the gaugino ones is reduced by a factor ~2 ~
with respect to the ones obtained for ~$M=10^7$ GeV. ~Since for 
low values of ~$\eta_K$, ~the overall properties are governed by the
value of ~$y$, ~the solutions for ~$y=0.55$ ~resemble those obtained for ~
$y=0.50$ ~at ~$M=10^7$ GeV. ~The bounds on the $CP$-odd mass necessary
for obtaining a light stop in the gauge-mediated scenario are somewhat
weaker, since there is a ~$\tan\beta$ ~enhancement associated with this
value, as is clear from eq. (\ref{eqn:a_mass}). Hence, for ~$y=0.55$, ~in 
the most restrictive case a 
light stop may only be obtained for $CP$-odd masses of the order of ~
2 TeV, ~while in the less restrictive case, masses of order ~900 GeV ~
are necessary. For ~$y = 0.35$, ~the bounds are raised to ~6 TeV ~and ~5 TeV, 
respectively.

\vskip 0.5cm

{\bf 6. CONCLUSIONS} 
\vskip 0.3cm

In this paper we have discussed the mapping to high energies of the low 
energy region of the MSSM parameter space, characterized by the presence
of a light stop and light chargino. The  parameter space considered is
consistent with present experimental constraints, such as precision tests,
the lower bound on the Higgs boson mass and ~$BR(b\rightarrow s\gamma)$ ~and 
therefore of special interest for physics at LEP2 and the Tevatron colliders.
For heavy top quark and small ~$\tan\beta$, ~the global pattern of the mapping
is determined by the proximity of the top-quark Yukawa coupling to its
IR fixed-point value and by the assumed scale at which supersymmetry
breaking is transmitted to the observable sector.
The general pattern of this mapping is the dominance
of the scalar masses (over the gaugino mass) in the supersymmetry breaking 
and the strong correlation between the soft SUSY breaking mass parameters ~
$m^2_Q$, ~$m^2_U$ ~and ~$m^2_{H_2}$. ~Moreover, we have identified
the conditions that are necessary for the spectrum to be consistent with a 
simple {\sl Ansatz} such as universality (or partial universality) of the
high energy values for the scalar masses. In particular, the ~$SO(10)$-type 
initial conditions, with universal left- and right-handed squark masses,
but with non-universal Higgs masses, are compatible with an interesting
subregion of the considered parameter space. In the scenario in which
the supersymmetry breaking is transferred to the observable sector via
gravitational interactions,
the conditions for such simple patterns are easier to satisfy away from
the IR fixed point. On the contrary, for low values of the messenger
masses, these patterns are easily obtained for small ~
$\tan\beta$ ~and values of the top quark Yukawa couplings associated with
large values of ~$h_t$ ~at ~$M_{GUT}$. ~In the latter
case the considered low energy spectra can be consistent with the typical
boundary conditions for the squark and gaugino appearing in 
gauge-mediated supersymmetry breaking. However, consistency with
with another generic prediction of those scenarios, ~$A_t\sim0$, ~
can be achieved only for extremely large values of the $CP$-odd Higgs
boson mass.
\vskip 1.0cm

\noindent {\bf Acknowledgements}

\noindent M.C. and C.W. would like to thank the Aspen Center for Physics,
where part of this work has been done.

\noindent P.H.Ch. would like to thank Prof. H. Haber and the Santa Cruz
Institute for Particle Physics for hospitality. His work and the
work of S.P. were partly 
supported by the joint U.S.--Polish Maria Sk\l odowska Curie grant.

\noindent M.O. thanks the INFN Sezione di Torino where part of this work has 
been done. His work was partly supported by the Deutsche 
Forschungsgemeinschaft under grant SFB--375/95 and 
by the European Commission TMR programmes ERBFMRX-CT96-0045
and ERBFMRX-CT96-0090.


\newpage
{\bf  APPENDIX}
\vskip 0.2cm

Here we collect formulae for various factors appearing in the solutions
to the RGEs. Solutions for the soft masses contain
the following functions of the gauge coupling constants:
\begin{eqnarray}
\xi_j(t) =
I\left[\sum_i \frac{M_i(0)}{M_{1/2}}\frac{a_j^i\alpha_i^2(t)}{\alpha_i(0)}
\right]
\end{eqnarray}
\begin{eqnarray}
\eta_K(t) =
I\left[\sum_i\frac{M_i^2(0)}{M_{1/2}^2}\frac{d_K^i\alpha_i^3(t)}{\alpha_i^2(0)}
\right]
\end{eqnarray}
\begin{eqnarray}
{\hat\xi}(t) =
H\left[\sum_i \frac{M_i(0)}{M_{1/2}}\frac{a^i_u\alpha_i^2(t)}{\alpha_i(0)}
\right]
\end{eqnarray}
\begin{eqnarray}
{\hat\eta}(t) =
H\left[\sum_i\frac{M_i^2(0)}{M_{1/2}^2}
\frac{\overline d^i\alpha_i^3(t)}{\alpha_i^2(0)}\right] 
+ H\left[\sum_i\frac{M_i(0)}{M_{1/2}}
\frac{\overline d^i\alpha_i^2(t)}{\alpha_i(0)}\xi_u(t)\right],
\end{eqnarray}
where the coefficients ~$a^i_j$ ~and ~$d^i_K$ ~
read: ~$a_u^i = (13/15, 3, 16/3)$, ~
$a_d^i = (7/15, 3, 16/3)$, ~$a_e^i = (3/5, 3, 0)$  ~($\xi_B = \xi_e$); ~
$d_Q^i = (1/15, 3, 16/3)$, ~$d_U^i = (16/15, 0, 16/3)$, ~
$d_D^i = (4/15, 0, 16/3)$, ~$d_L^i = d^i_{H_1} = d^i_{H_2} = (3/5, 3, 0)$, ~
$d_E^i = (12/5, 0, 0)$; ~
$\overline d^i\equiv d^i_Q + d^i_U + d^i_{H_2}$ ~and ~$I$, ~and ~$H$ ~
are defined by:
\begin{eqnarray}
I[f(t)] = \int_0^t f(t^{\prime})dt^{\prime}
\end{eqnarray}
\begin{eqnarray}
H[f(t)] = \int_0^t f(t^{\prime}) dt^{\prime}
- \frac{1}{F(t)}\int_0^t F(t^{\prime})f(t^{\prime})dt^{\prime}
\end{eqnarray}
with ~$F(t)$ ~given by eqs. (\ref{eqn:fft}).
The factors ~$M_i(0)/M_{1/2}\neq1$ ~appear because we do not assume exact
gauge coupling unification and, by convention, ~$M_{1/2}\equiv M_3(0)$. ~
Parameters ~$\xi_i$ ~and ~$\eta_K$ ~can be computed analytically.
Parameters ~$\hat\xi$ ~and ~$\hat\eta$ ~require numerical integration.
We give below (using the assumption (\ref{eqn:unass}))
their typical values for few scales ~$M$ ~and three values of ~$\alpha_3$. ~
For other values of ~~$\alpha_3$, ~$\hat\xi$ ~and ~$\hat\eta$ ~can be easily 
obtained by interpolation.

\vskip 0.3cm 
{\bf Table 1.}  Typical ~$\hat\xi$ ~values
\begin{center}
\begin{tabular}{|| c | c || c | c | c | c | c | c ||} \hline
\multicolumn{2}{||l||}{~$\hat\xi$}&\multicolumn{6}{c||}{$\,M$ [GeV]}\\
\cline{3-8}
\multicolumn{2}{||l||}{}&
    $4\times10^{16}$ & $2\times10^{16}$ & $1\times10^{16}$ &
    $1\times10^{10}$ & $1\times10^7$    & $1\times10^5$\\ 
\hline\hline
&   0.115   &     2.22      &    2.16       &   2.10        &
                  1.08      &   0.640       &  0.374        \\ 
\cline{2-8}$\alpha_3$ 
&   0.120   &     2.29      &    2.23       &   2.17        &
                  1.11      &   0.663       &  0.388        \\ 
\cline{2-8}
&   0.125   &     2.36      &    2.30       &   2.23        &
                  1.15      &   0.686       &  0.402        \\ 
\hline
\end{tabular}
\end{center}

\newpage
{\bf Table 2.}  Typical ~$\hat\eta$ ~values
\begin{center}
\begin{tabular}{|| c | c || c | c | c | c | c | c ||} \hline
\multicolumn{2}{||l||}{~$\hat\eta$}&\multicolumn{6}{c||}{$\,M$ [GeV]}\\
\cline{3-8}
\multicolumn{2}{||l||}{}&
    $4\times10^{16}$ & $2\times10^{16}$ & $1\times10^{16}$ &
    $1\times10^{10}$ & $1\times10^7$    & $1\times10^5$\\ 
\hline\hline
&   0.115   &     12.7      &    12.1       &   11.6        &
                  4.16      &    1.10       &   0.987        \\ 
\cline{2-8}$\alpha_3$ 
&   0.120   &     13.4      &    12.8       &   12.3        &
                  4.39      &    2.10       &   1.04        \\ 
\cline{2-8}
&   0.125   &     14.1      &    13.5       &   12.9        &
                  4.63      &    2.22       &   1.09        \\ 
\hline
\end{tabular}
\end{center}

\vskip 0.3cm 
{\bf Table 3.}  Typical ~$\xi_u$ ~values
\begin{center}
\begin{tabular}{|| c | c || c | c | c | c | c | c ||} \hline
\multicolumn{2}{||l||}{~$\xi_u$}&\multicolumn{6}{c||}{$\,M$ [GeV]}\\
\cline{3-8}
\multicolumn{2}{||l||}{}&
    $4\times10^{16}$ & $2\times10^{16}$ & $1\times10^{16}$ &
    $1\times10^{10}$ & $1\times10^7$    & $1\times10^5$\\ 
\hline\hline
&   0.115   &     3.94      &    3.84       &   3.75        &
                  2.03      &    1.24       &  0.736        \\ 
\cline{2-8}$\alpha_3$ 
&   0.120   &     4.07      &    3.97       &   3.88        &
                  2.10      &    1.29       &  0.764        \\ 
\cline{2-8}
&   0.125   &     4.20      &    4.10       &   4.01        &
                  2.17      &    1.33       &  0.792        \\ 
\hline
\end{tabular}
\end{center}

\vskip 0.3cm 
{\bf Table 4.}  Typical ~$\xi_B$ ~values
\begin{center}
\begin{tabular}{|| c | c || c | c | c | c | c | c ||} \hline
\multicolumn{2}{||l||}{~$\xi_B$}&\multicolumn{6}{c||}{$\,M$ [GeV]}\\
\cline{3-8}
\multicolumn{2}{||l||}{}&
    $4\times10^{16}$ & $2\times10^{16}$ & $1\times10^{16}$ &
    $1\times10^{10}$ & $1\times10^7$    & $1\times10^5$\\ 
\hline\hline
&   0.115   &    0.614      &   0.590       &  0.566        &
                 0.210      &   0.102       &  0.050        \\ 
\cline{2-8}$\alpha_3$ 
&   0.120   &    0.605      &   0.581       &  0.557        &
                 0.206      &   0.099       &  0.049        \\ 
\cline{2-8}
&   0.125   &    0.597      &   0.573       &  0.549        &
                 0.202      &   0.097       &  0.047        \\ 
\hline
\end{tabular}
\end{center}

\vskip 0.3cm 
{\bf Table 5.}  Typical ~$\eta_Q$ ~values
\begin{center}
\begin{tabular}{|| c | c || c | c | c | c | c | c ||} \hline
\multicolumn{2}{||l||}{~$\eta_Q$}&\multicolumn{6}{c||}{$\,M$ [GeV]}\\
\cline{3-8}
\multicolumn{2}{||l||}{}&
    $4\times10^{16}$ & $2\times10^{16}$ & $1\times10^{16}$ &
    $1\times10^{10}$ & $1\times10^7$    & $1\times10^5$\\ 
\hline\hline
&   0.115   &     6.87      &    6.65       &   6.44        &
                  2.85      &    1.54       &   0.836       \\ 
\cline{2-8}$\alpha_3$ 
&   0.120   &     7.27      &    7.04       &   6.81        &
                  3.01      &    1.62       &   0.876       \\ 
\cline{2-8}
&   0.125   &     7.69      &    7.44       &   7.19        &
                  3.17      &    1.70       &   0.917       \\ 
\hline
\end{tabular}
\end{center}

\vskip 0.3cm 
{\bf Table 6.}  Typical ~$\eta_U$ ~values
\begin{center}
\begin{tabular}{|| c | c || c | c | c | c | c | c ||} \hline
\multicolumn{2}{||l||}{~$\eta_U$}&\multicolumn{6}{c||}{$\,M$ [GeV]}\\
\cline{3-8}
\multicolumn{2}{||l||}{}&
    $4\times10^{16}$ & $2\times10^{16}$ & $1\times10^{16}$ &
    $1\times10^{10}$ & $1\times10^7$    & $1\times10^5$\\ 
\hline\hline
&   0.115   &     6.42      &    6.22       &   6.03        &
                  2.74      &    1.50       &  0.817        \\ 
\cline{2-8}$\alpha_3$ 
&   0.120   &     6.84      &    6.62       &   6.41        &
                  2.90      &    1.58       &  0.858        \\ 
\cline{2-8}
&   0.125   &     7.26      &    7.03       &   6.81        &
                  3.06      &    1.66       &  0.900        \\ 
\hline
\end{tabular}
\end{center}

\vskip 0.3cm 
{\bf Table 7.}  Typical ~$\eta_{H_2}$ ~values
\begin{center}
\begin{tabular}{|| c | c || c | c | c | c | c | c ||} \hline
\multicolumn{2}{||l||}{~$\eta_{H_2}$}&\multicolumn{6}{c||}{$\,M$ [GeV]}\\
\cline{3-8}
\multicolumn{2}{||l||}{}&
    $4\times10^{16}$ & $2\times10^{16}$ & $1\times10^{16}$ &
    $1\times10^{10}$ & $1\times10^7$    & $1\times10^5$\\ 
\hline\hline
&   0.115   &     0.561    &    0.529     &   0.499      &
                  0.128    &    0.049     &   0.020      \\ 
\cline{2-8}$\alpha_3$ 
&   0.120   &     0.544    &    0.513     &   0.484      &
                  0.123    &    0.047     &   0.019      \\ 
\cline{2-8}
&   0.125   &     0.530    &    0.499     &   0.471      &
                  0.118    &    0.045     &   0.018      \\ 
\hline
\end{tabular}
\end{center}

\newpage
\setcounter{figure}{0}
\begin{figure}
\psfig{figure=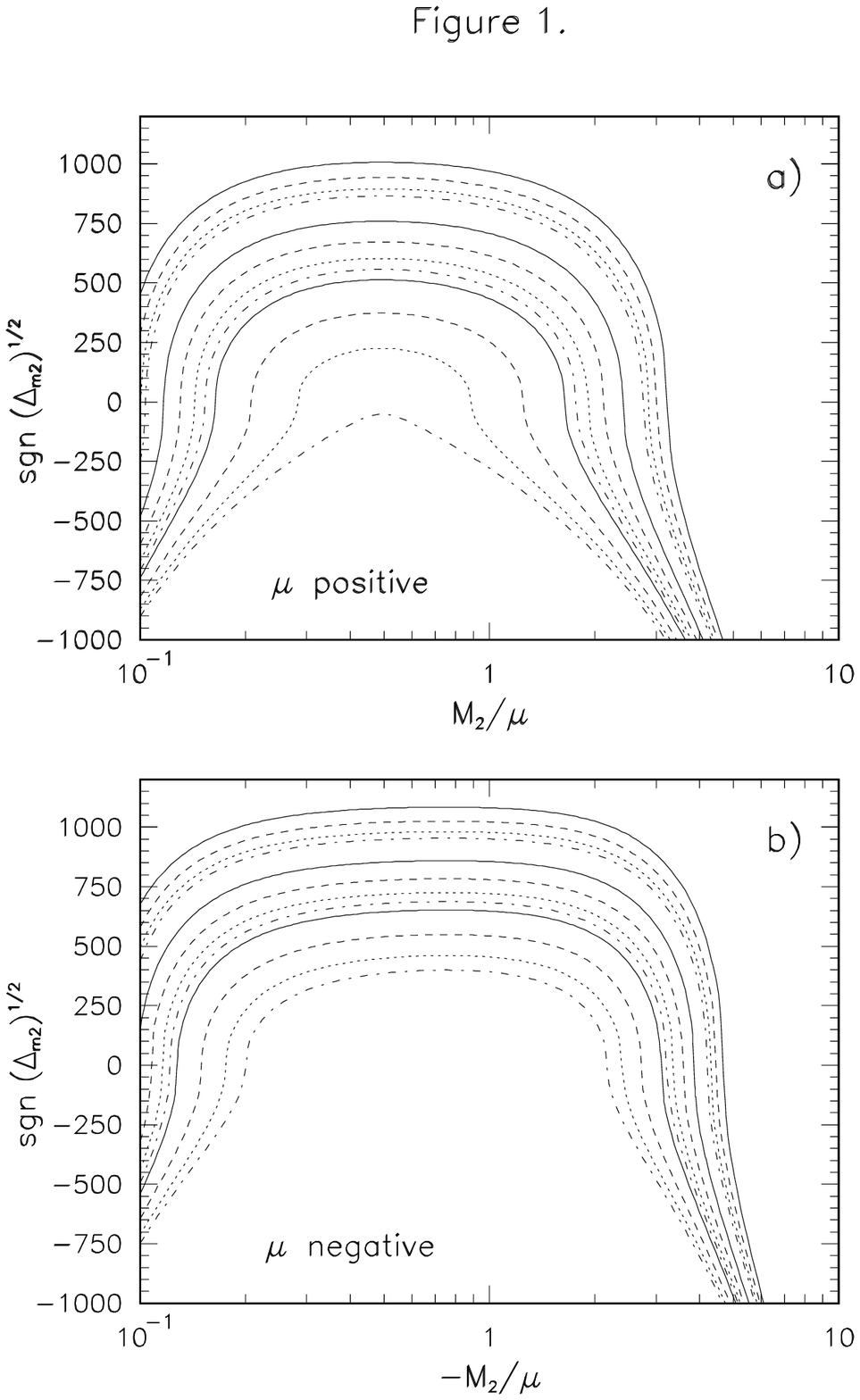,width=16.0cm,height=18.0cm}
\caption{$\Delta_{m^2}$ defined in eqs. 
(\ref{eqn:sumrule}),(\ref{eqn:sumr_phys})
as a function of $r\equiv M_2/|\mu|$ ~for ~$M_A=1000$ (solid lines), ~
$750$ (dashed), $500$ (dotted) and $250$ (dash-dotted) GeV. ~Different lines
of the same type correspond to $M_{\tilde t_1}=1000$ (upper), $750$ 
(middle) and $500$ (lower) GeV. Cases with positive and negative $\mu$ 
are shown in a) and b) respectively. 
Values of ~$m_t=175$ GeV, ~$\tan\beta=1.6$, ~
$m_{C^+}=90$ GeV ~and ~$M_{\tilde t_2}=60$ GeV are taken.}
\label{fig1}
\end{figure}

\newpage
\setcounter{figure}{1}
\begin{figure}
\psfig{figure=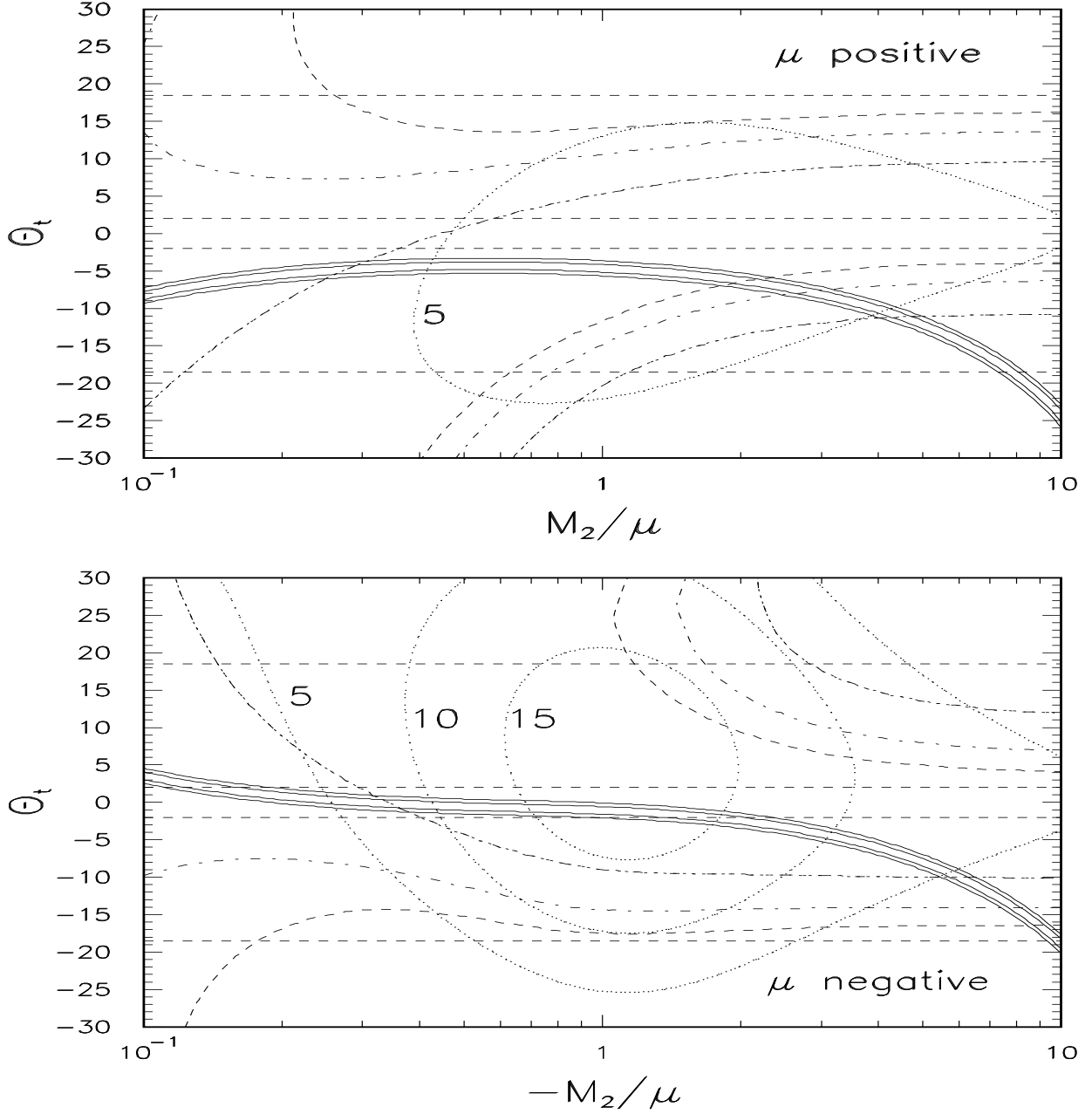,width=18cm,height=19.0cm}
\caption{{\bf a.} ~Contours of $\Delta_A=50$ (internal solid lines) and 
100 GeV(external ones) in the plane $(r, ~\theta_{\tilde t})$ ~for 
$M_{\tilde t_1}=1000$ GeV (for positive
and negative values of ~$\mu$ ~in the upper and lower panels respectively).
Regions excluded by ~$M_h>60$ GeV (horizontal lines) ~and ~$0.98\times10^{-4} 
< BR(b\rightarrow s\gamma)<3.66\times10^{-4}$ ~for ~$M_A=1000$, 500 and
250 GeV  ~are marked by dashed, dot-dashed, and dot-dot-dashed lines
respectively (contours of $\Delta_A$ are ~$M_A$ independent, see eq.
(\ref{eqn:sumr_phys})). Dotted contours show values of ~
$\delta R_b\times10^4$. ~Other parameters as in Fig. 1.}
\label{fig2a}
\end{figure}

\newpage
\setcounter{figure}{1}
\begin{figure}
\psfig{figure=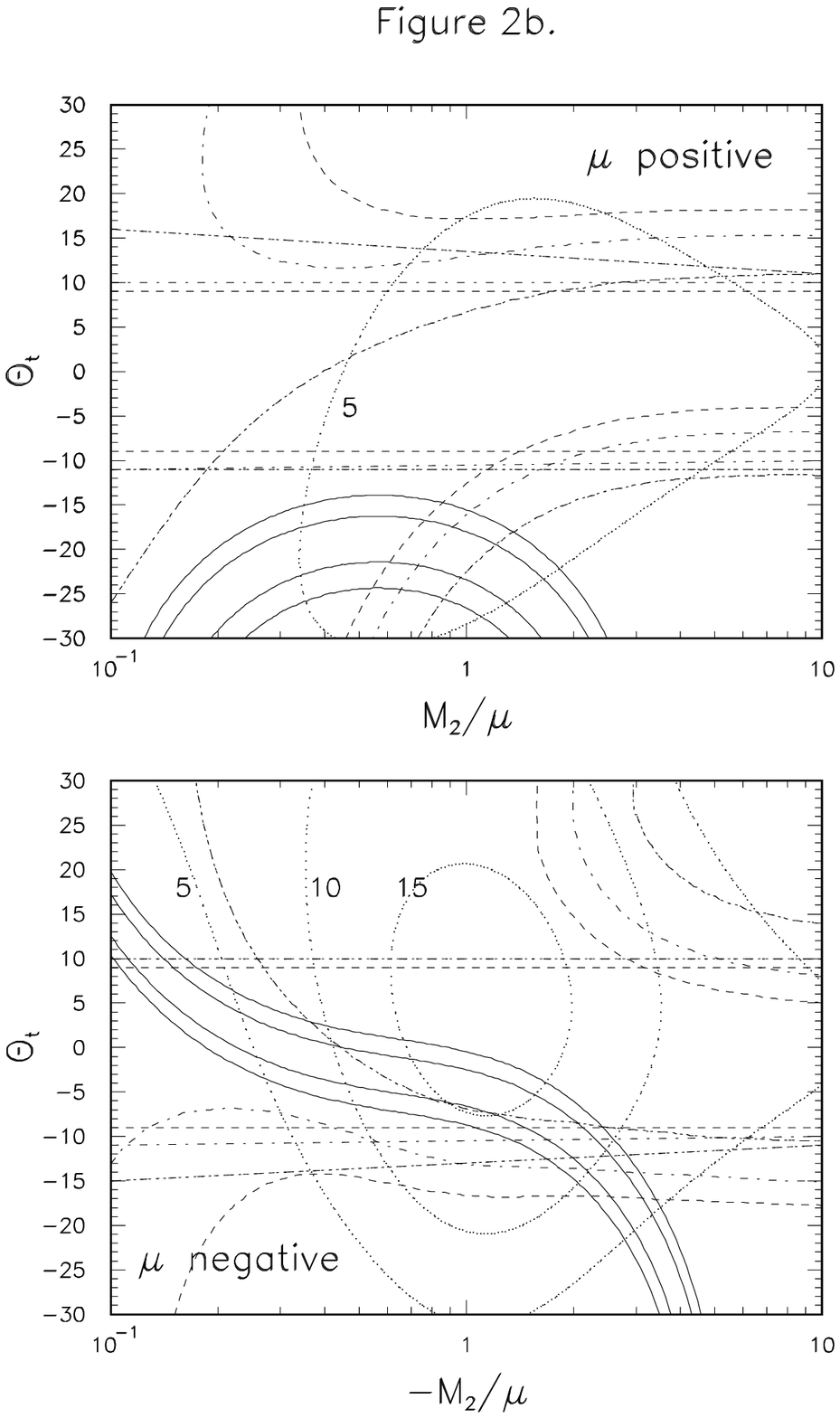,width=15cm,height=19.0cm}
\caption{{\bf b.} ~As in Fig. 2a, but for $M_{\tilde t_1}=500$ GeV.}
\label{fig2b}
\end{figure}

\newpage
\setcounter{figure}{2}
\begin{figure}
\psfig{figure=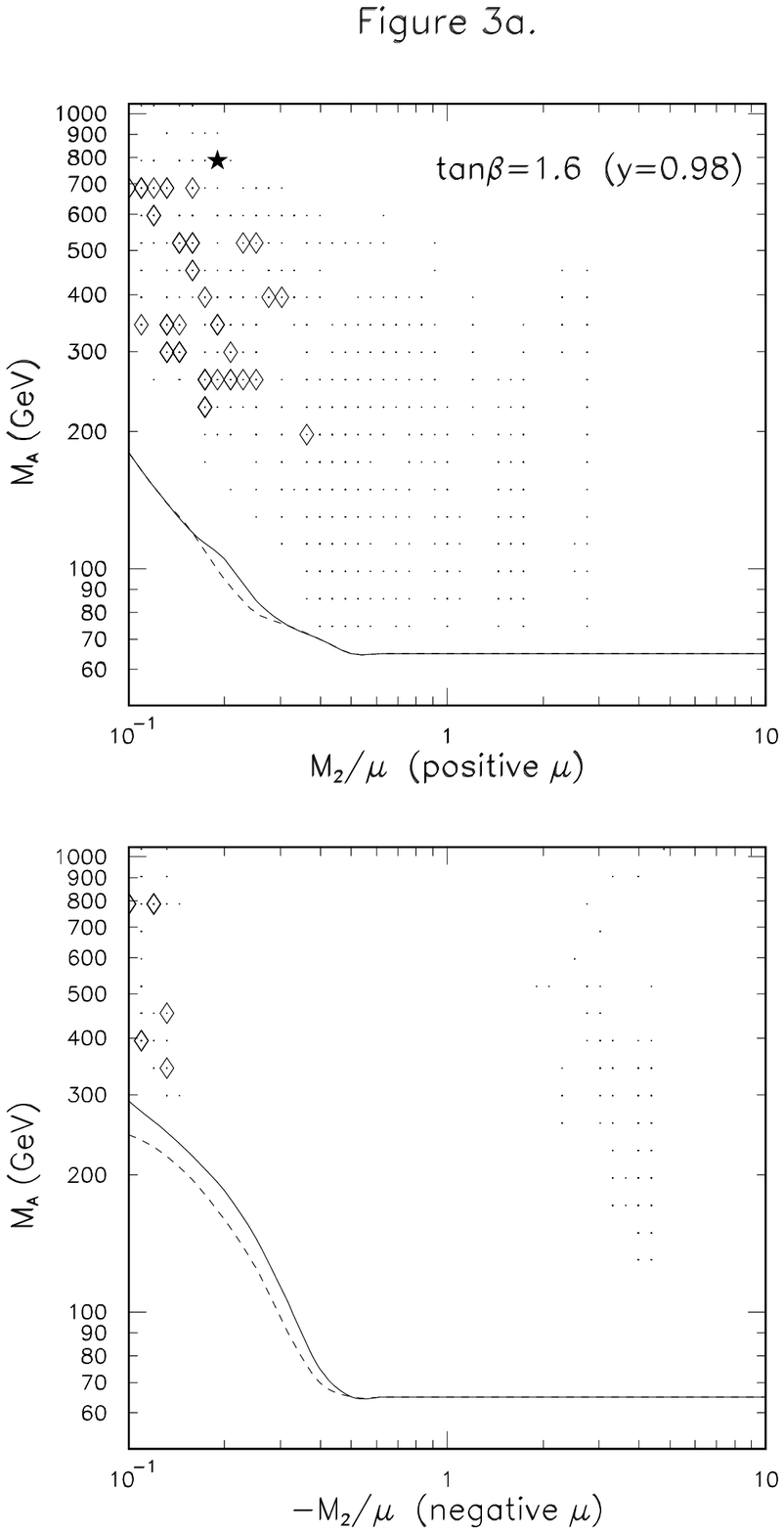,width=16cm,height=17.0cm}
\caption{{\bf a.} ~Regions in the ~$(r, M_A)$ ~
plane allowed by experimental constraints listed in the text (for positive
and negative values of ~$\mu$ ~in the upper and lower panels respectively)
for ~$\tan\beta=1.6$ ~(corresponding for ~
$M_{GUT}=2\times10^{16}$ GeV ~to ~$y=0.98$). ~
Regions above the dashed lines are allowed by ~$M_h>60$ GeV ~and ~
$BR(b\rightarrow s\gamma)$. ~In regions above the solid lines,
in addition, ~$\chi^2 < \chi^2_{best ~SM} + 2$ (see the text). The
black star shows the point that can be obtained with fully universal boundary
conditions at ~$M_{GUT}$ (i.e. with $m^2_Q(0)$, $m^2_U(0)$, $m^2_{H_1}(0)$ 
and $m^2_{H_2}$ ~all equal within 10\%). The white rhombs denote points that
can be realized with $m^2_Q(0)$, $m^2_U(0)$, $m^2_{H_2}(0)$ equal and the
dots correspond to points that require $SO(10)$-type initial values 
(i.e. $m^2_Q(0)=m^2_U(0)$ within 10\%). All marked points are consistent
with our experimental cuts. Values of the other parameters as in Fig. 1.}
\label{fig3a}
\end{figure}

\newpage
\setcounter{figure}{2}
\begin{figure}
\psfig{figure=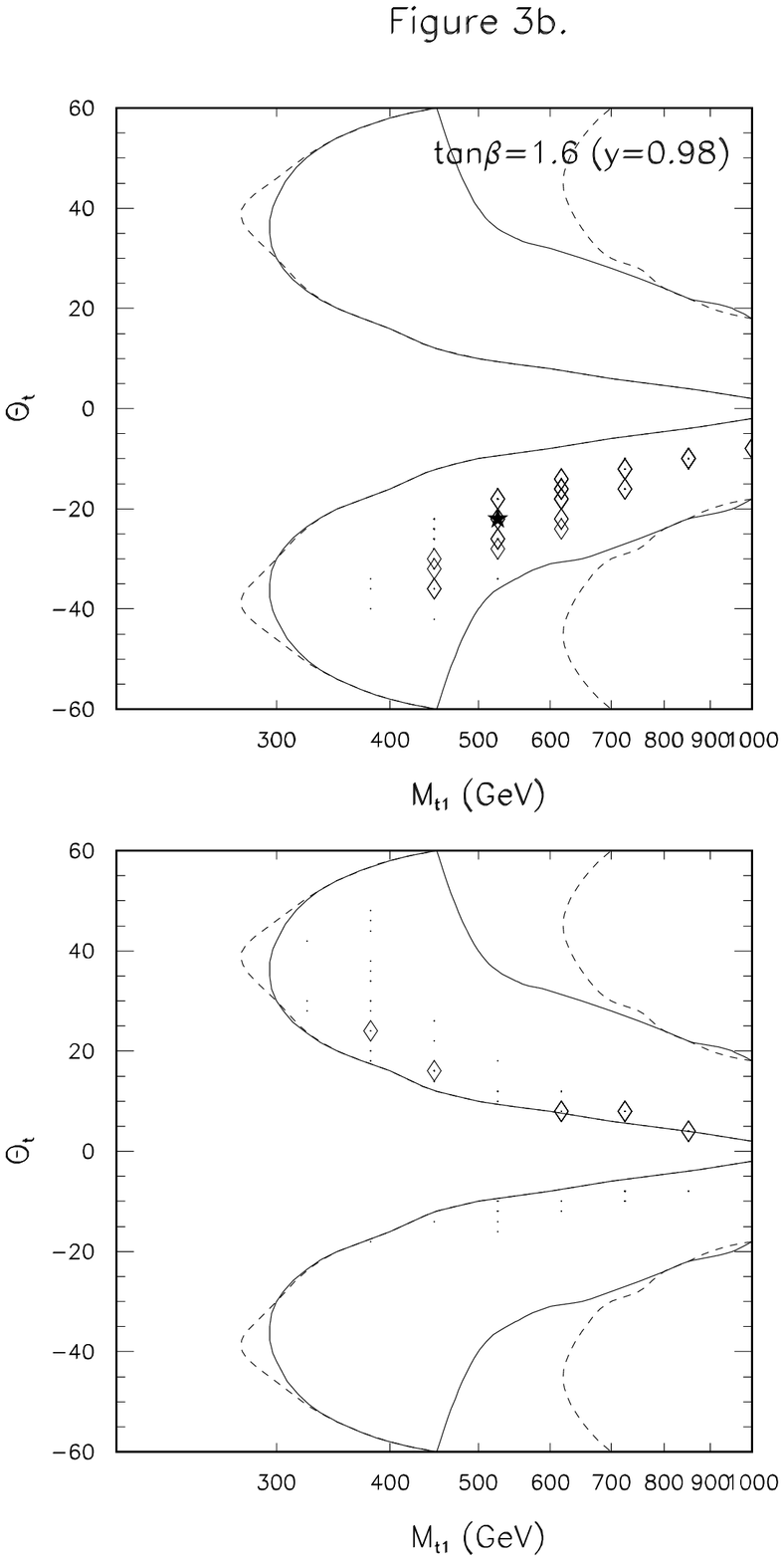,width=16cm,height=17.0cm}
\caption{{\bf b.} ~As in Fig. 3a, but in the ~$(M_{\tilde t_1}, ~
\theta_{\tilde t})$ plane. Allowed regions are inside marked contours.}
\label{fig3b}
\end{figure}

\newpage
\setcounter{figure}{3}
\begin{figure}
\psfig{figure=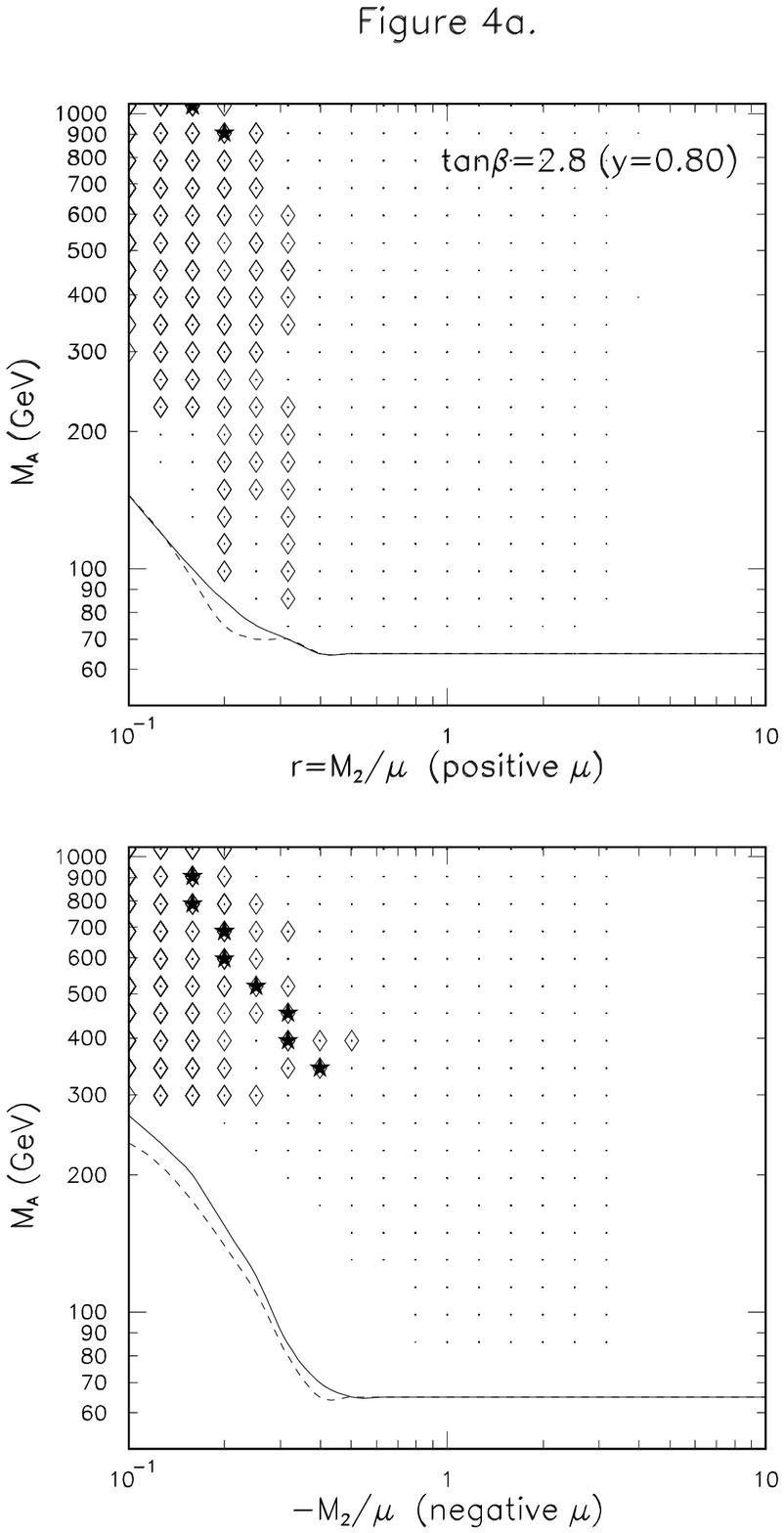,width=16cm,height=17.0cm}
\caption{{\bf a.} ~As in Fig. 3a, ~but for ~$\tan\beta=2.8$ ~(corresponding 
for ~$M_{GUT}=2\times10^{16}$ GeV ~to ~$y=0.80$).}
\label{fig4a}
\end{figure}

\newpage
\setcounter{figure}{3}
\begin{figure}
\psfig{figure=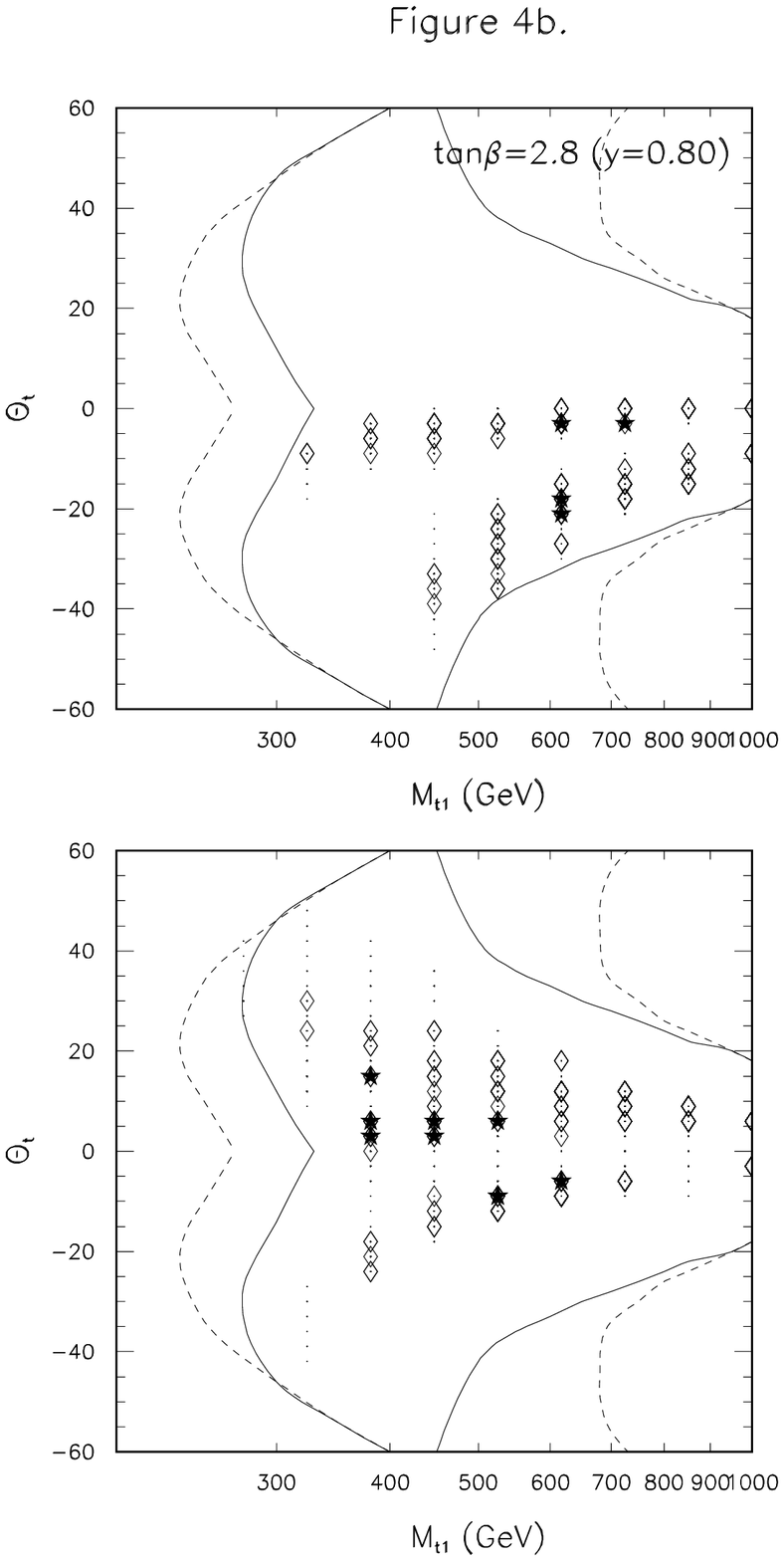,width=16cm,height=17.0cm}
\caption{{\bf b.} ~As in Fig. 4a, but in the ~
$(M_{\tilde t_1}, ~\theta_{\tilde t})$ plane.}
\label{fig4b}
\end{figure}

\newpage
\setcounter{figure}{4}
\begin{figure}
\psfig{figure=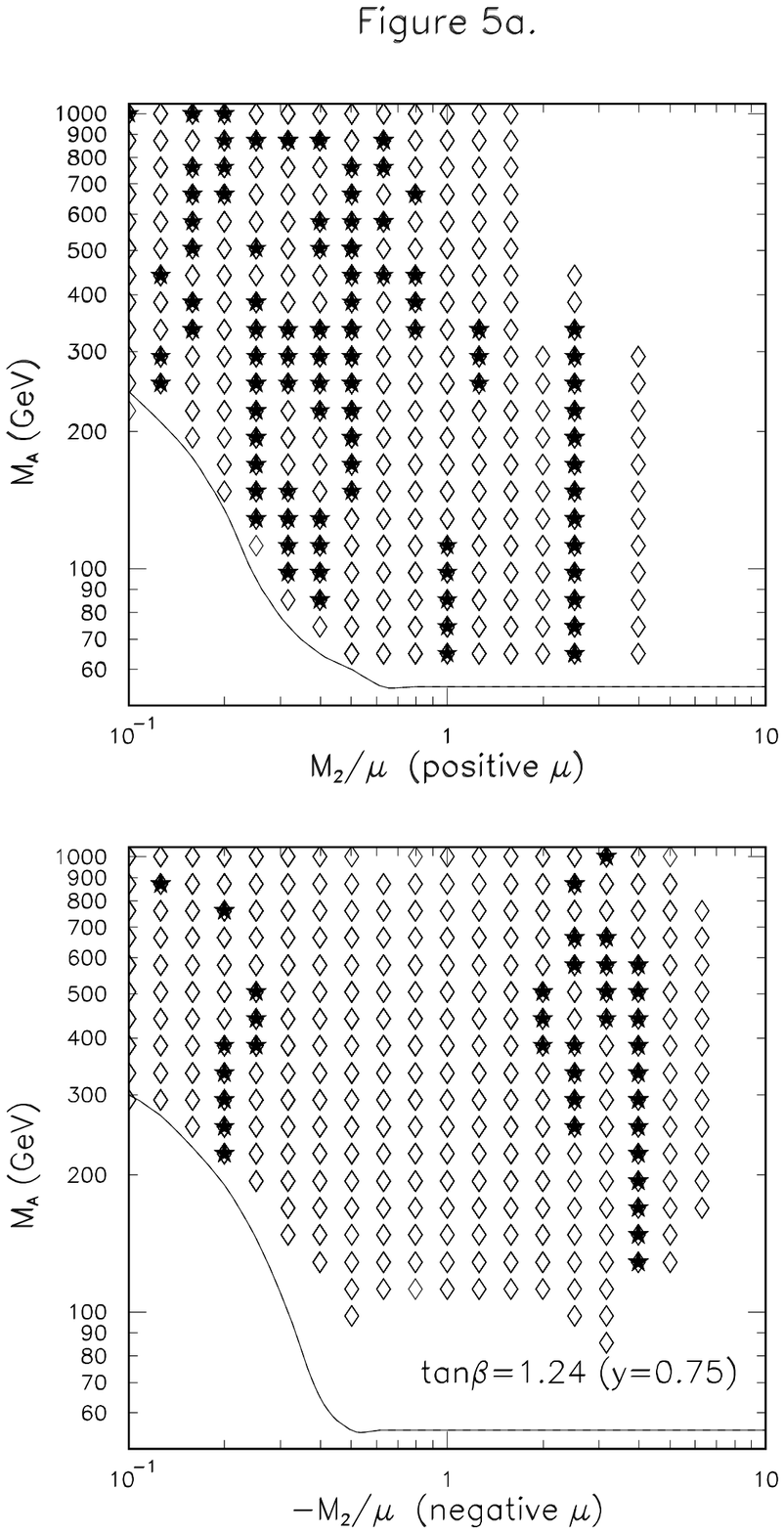,width=16cm,height=17.0cm}
\caption{{\bf a.} ~As in Fig. 3a, ~but for ~$\tan\beta=1.2$ ~(corresponding, 
for ~$M=10^7$ GeV, ~to ~$y=0.75$). The black stars (white rhombs) show the 
points that can be obtained with more (less) restrictive boundary conditions
at ~$M=10^7$ GeV (see the text for explanation).}
\label{fig5a}
\end{figure}

\newpage
\setcounter{figure}{4}
\begin{figure}
\psfig{figure=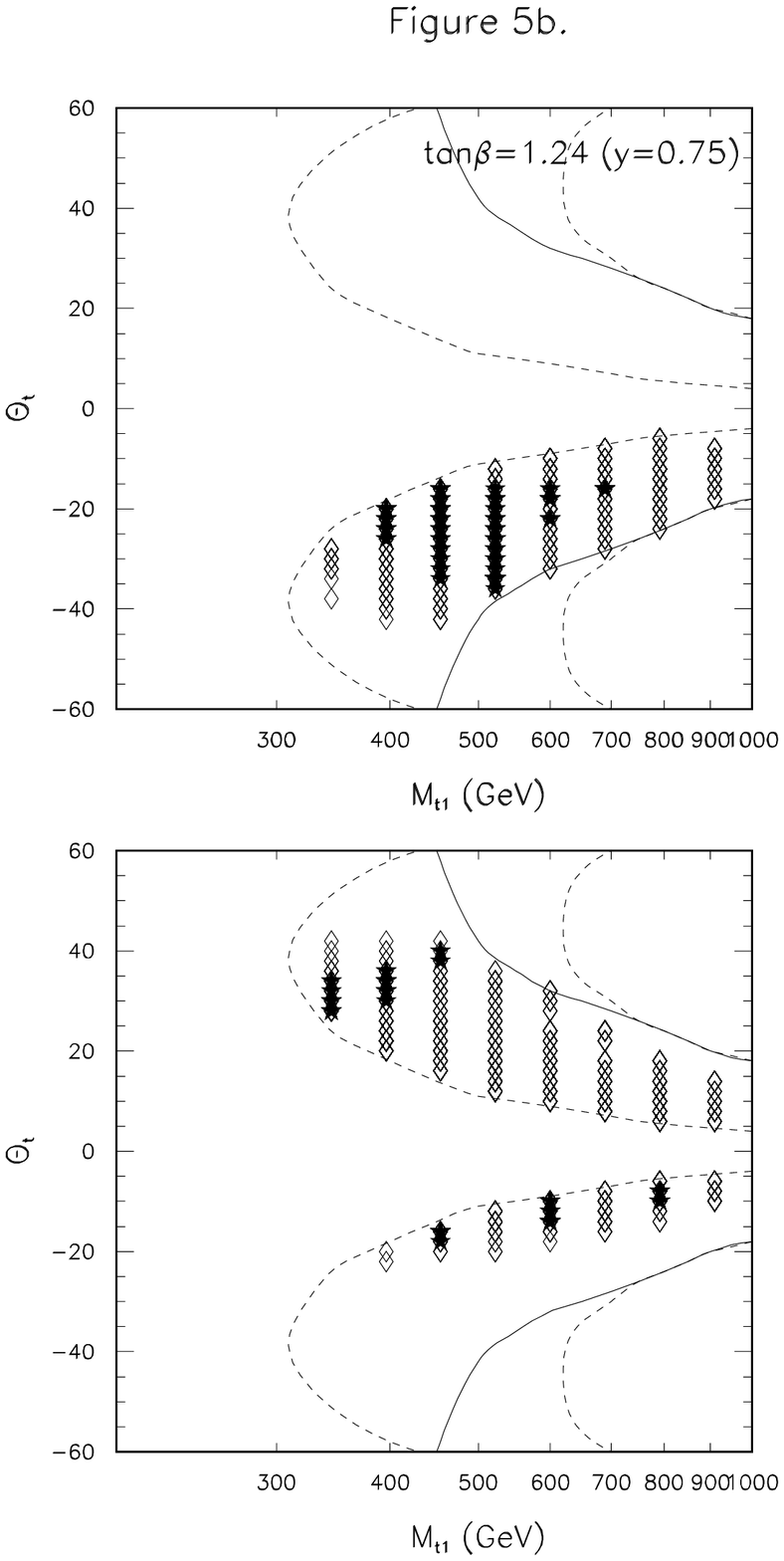,width=16cm,height=17.0cm}
\caption{{\bf b.} ~As in Fig. 5a, but in the ~$(M_{\tilde t_1}, ~
\theta_{\tilde t})$ plane.}
\label{fig5b}
\end{figure}

\newpage
\setcounter{figure}{5}
\begin{figure}
\psfig{figure=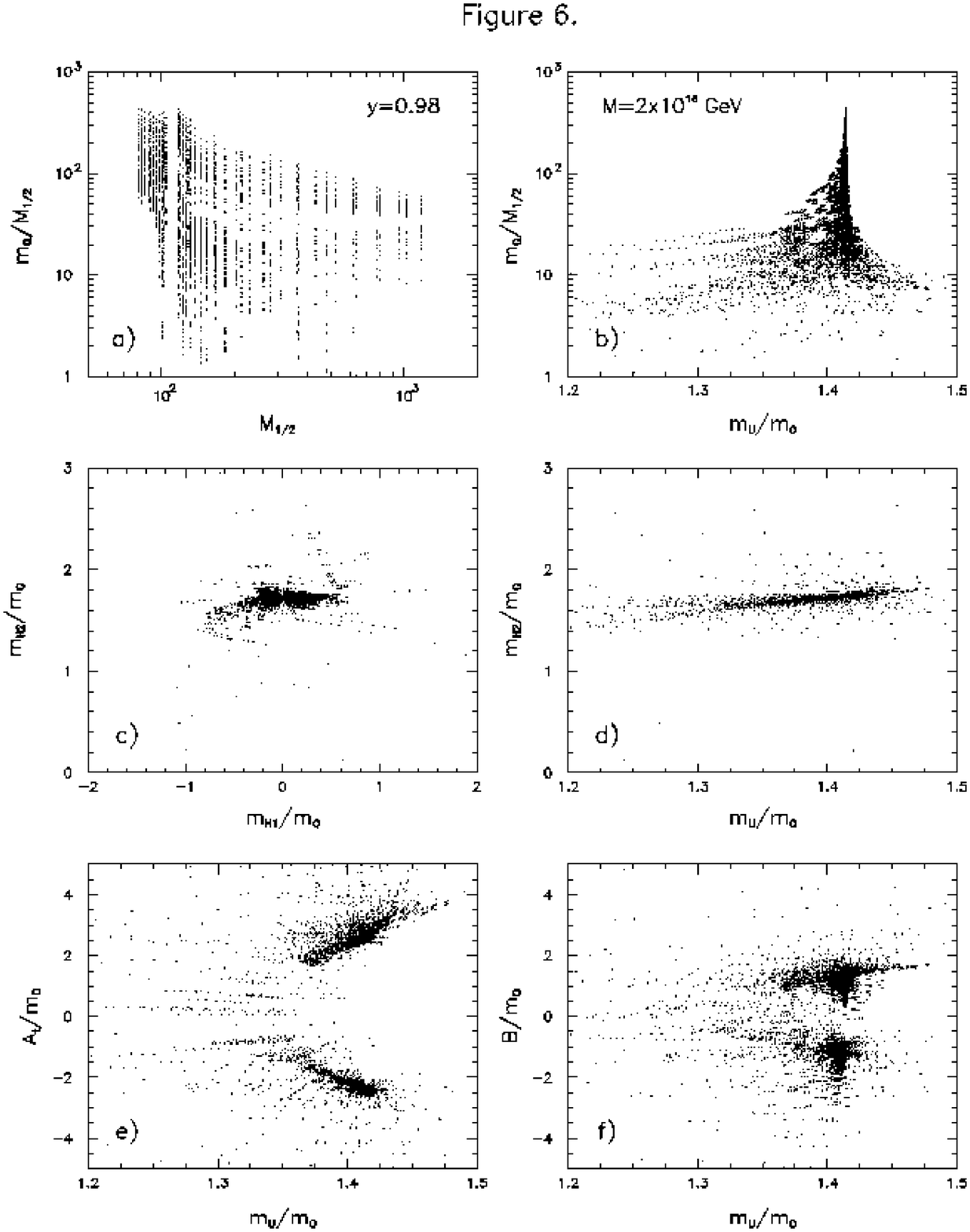,width=16cm,height=17.0cm}
\caption{Soft supersymmetry breaking parameters at the scale 
$M = M_{GUT}=2\times10^{16}$ GeV,
for $y_{GUT} = 0.98$ obtained by mapping the low-energy parameter space
subject to the conditions specified in the section 4.}
\label{fig6}
\end{figure}

\newpage
\setcounter{figure}{6}
\begin{figure}
\psfig{figure=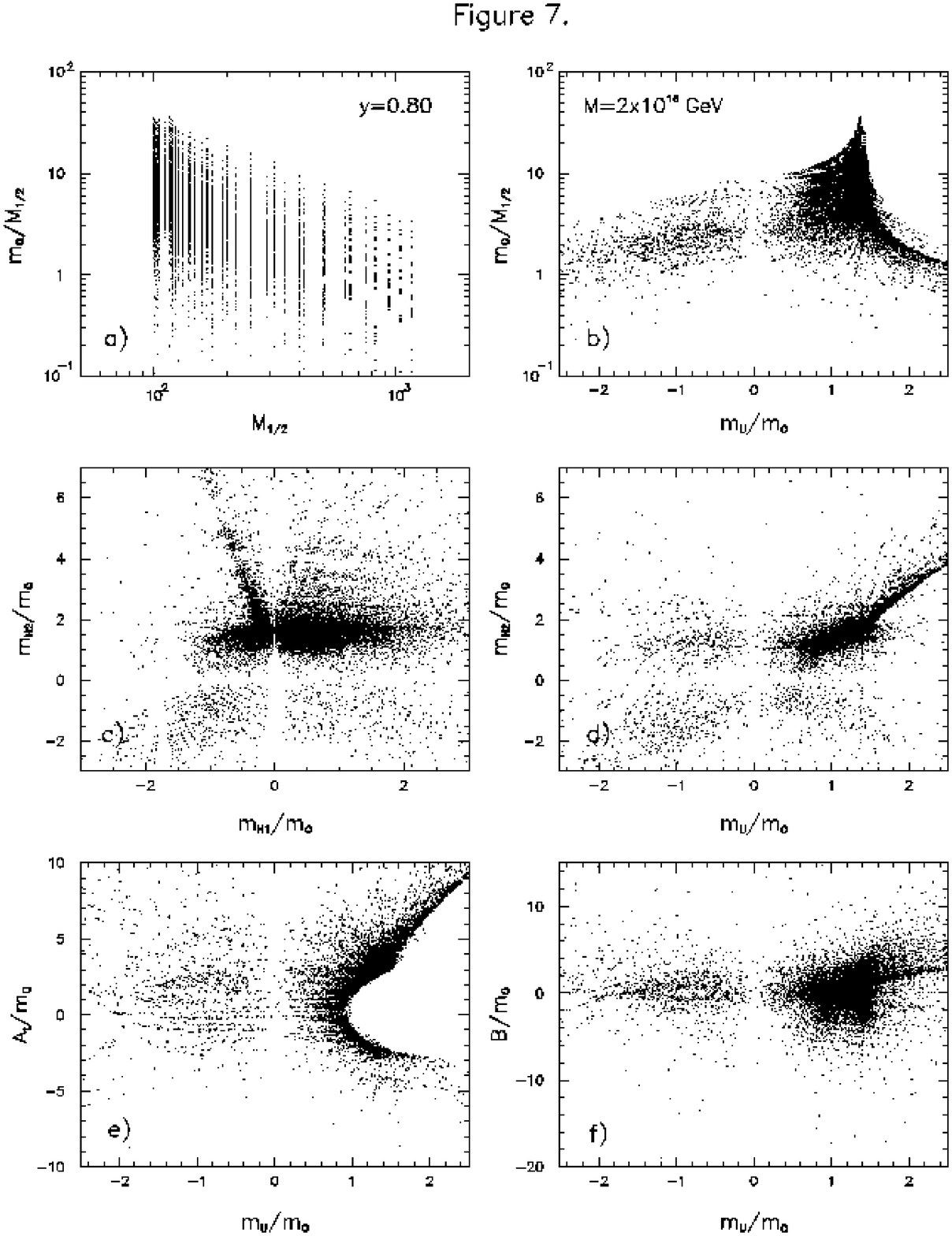,width=16cm,height=17.0cm}
\caption{As in Fig. 6, but for $y_{GUT} = 0.8$.}
\label{fig7}
\end{figure}

\newpage
\setcounter{figure}{7}
\begin{figure}
\psfig{figure=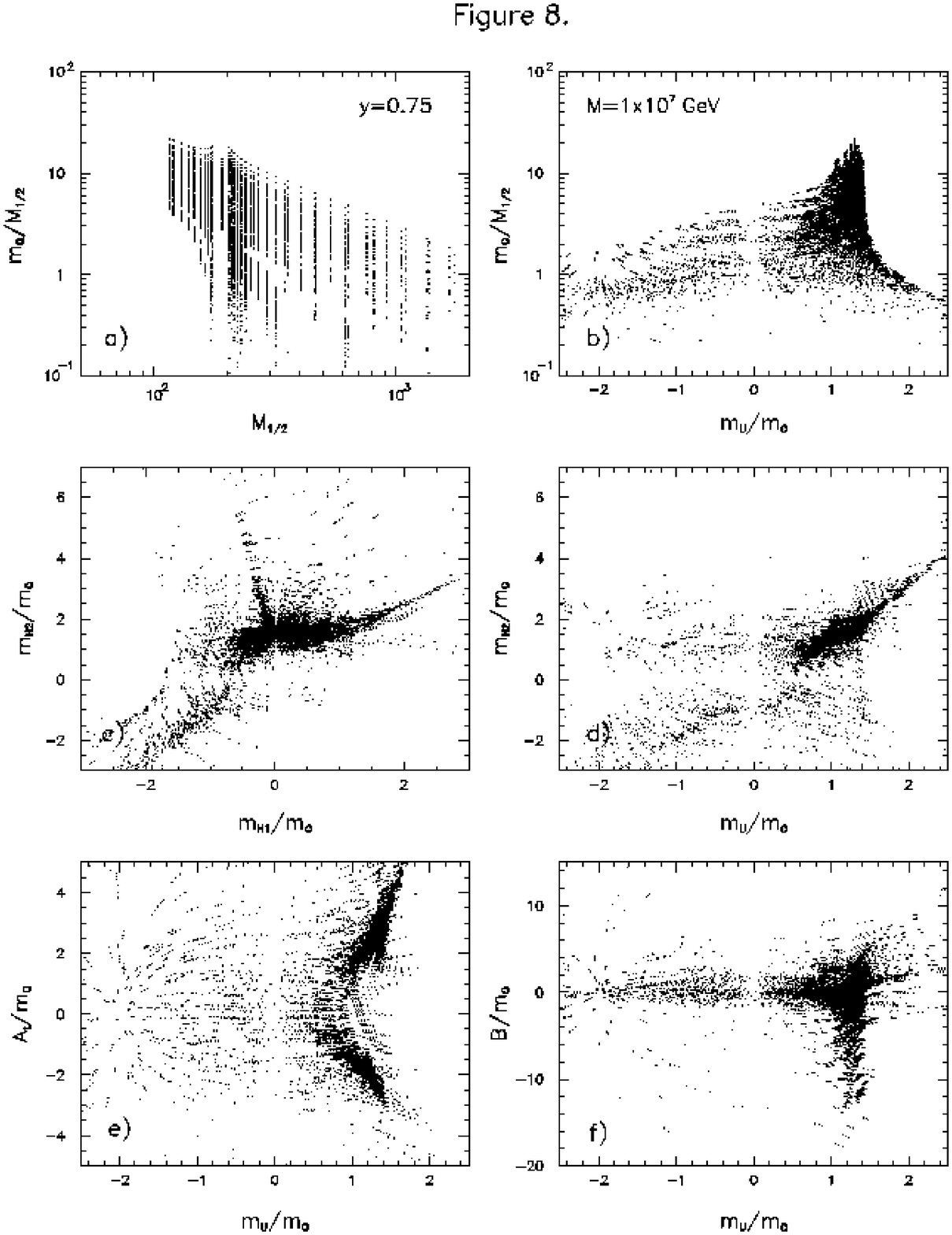,width=16cm,height=17.0cm}
\caption{As in Fig. 6, but for $M = 10^7$ GeV and $y(t) = 0.75$.}
\label{fig8}
\end{figure}

\newpage
\setcounter{figure}{8}
\begin{figure}
\psfig{figure=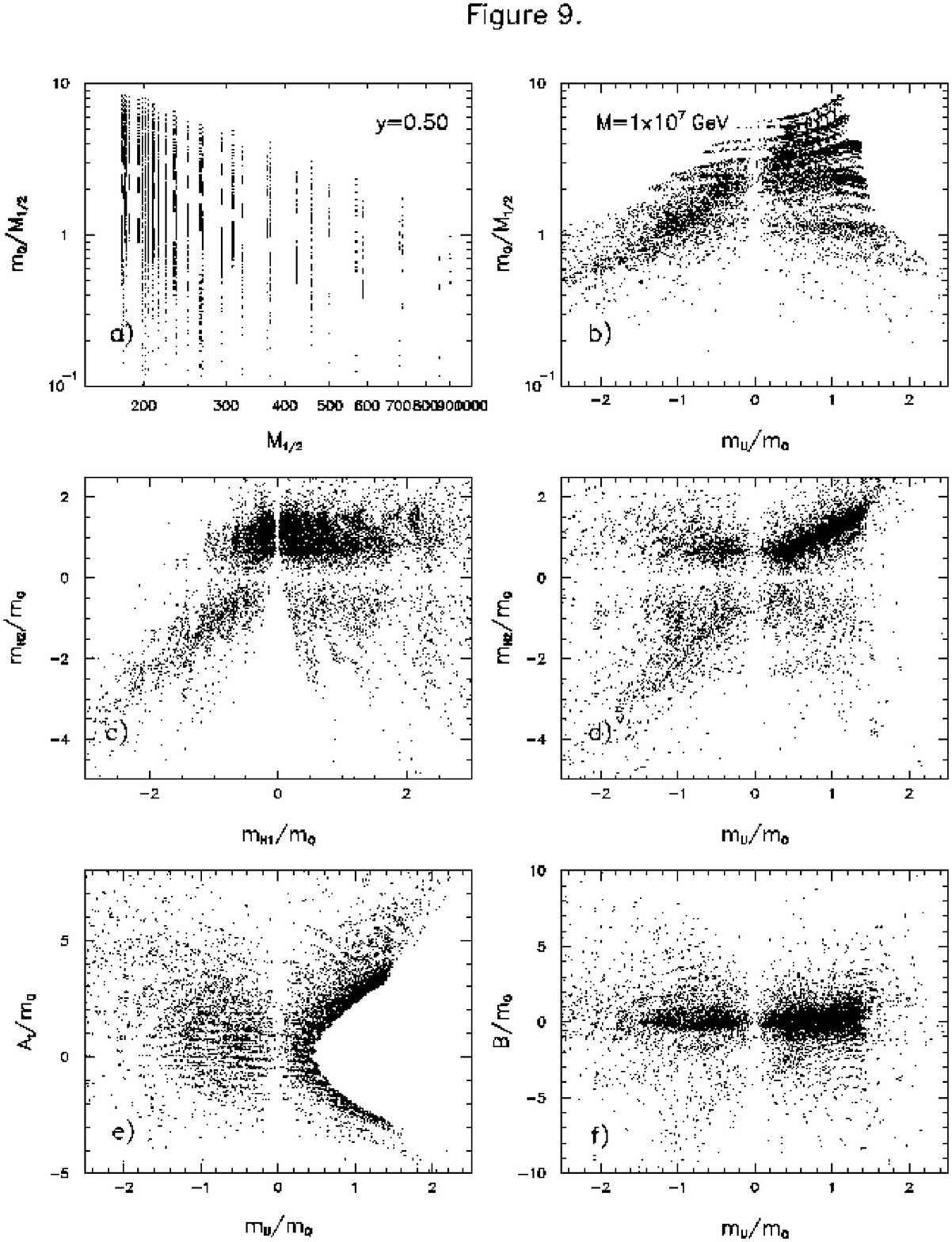,width=16cm,height=17.0cm}
\caption{As in Fig. 8, but for $y(t) = 0.50$.}
\label{fig9}
\end{figure}

\newpage
\setcounter{figure}{9}
\begin{figure}
\psfig{figure=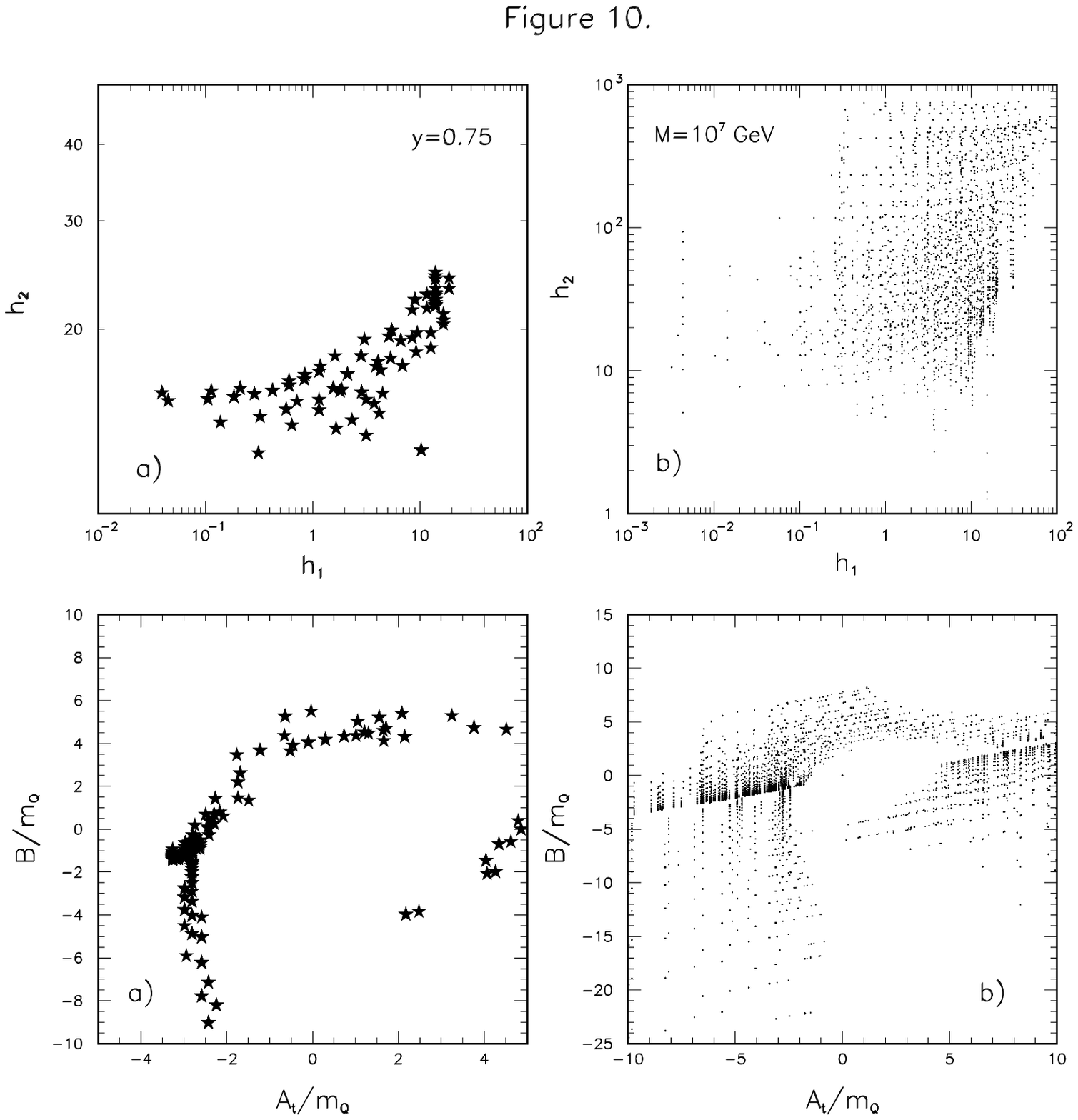,width=15cm,height=19.0cm}
\caption{Ratio ($h_{1,2}$) of the squared Higgs boson mass parameters, 
$m_{H_{1,2}}^2(0)$, to the ones predicted in gauge-mediated supersymmetry 
breaking models, for the case of minimal relations between the
squark and gaugino masses (left) and for the unrestricted case (right).
The behaviour of the soft supersymmetry breaking parameters
$A_t$ and $B$ at the scale $M=10^7$ GeV is also shown here.}
\label{fig10}
\end{figure}

\end{document}